\documentclass[useAMS,usenatbib,usegraphicx]{mn2e}

\title{The radial structure of galaxy groups and clusters}
\author[Y.~Ascasibar et al.]
{Y.~Ascasibar,$^{1, 2, 3}$ G.~Yepes,$^2$ V.~M\"uller$^3$ and S.~Gottl\"ober$^3$\\
 $^1$Theoretical Physics, 1Keble Road, Oxford OX1 3NP (United Kingdom)\\
 $^2$Grupo de Astrof\'\i sica, Universidad Aut\'onoma de Madrid, Madrid E-28049 (Spain)\\
 $^3$Astrophysikalisches Institut Potsdam, An der Sternwarte 16, Potsdam D-14482 (Germany)
}

\newcommand{\rs}{r_{\rm s}}
\newcommand{\rhos}{\rho_{\rm s}}
\newcommand{\ri}{r_{\rm i}}
\newcommand{\rhoi}{\rho_{\rm i}}
\newcommand{\ax}{a_{\rm x}}
\newcommand{\Lx}{L_{\rm X}}
\newcommand{\Tx}{T_{\rm X}}
\newcommand{\rhodm}{\rho_{\rm dm}}
\newcommand{\rhogas}{\rho_{\rm g}}
\newcommand{\dd}{{\rm d}}

\newcommand{\be}{\begin{equation}}
\newcommand{\ee}{\end{equation}}
\newcommand{\bea}{\begin{eqnarray}}
\newcommand{\eea}{\end{eqnarray}}

\newcommand{\Msun}{M$_\odot$}
\newcommand{\msun}{\Msun\ }

\newcommand{\G}{{\sc Gadget}}
\newcommand{\g}{\G\ }

\newcommand{\LCDM}{$\Lambda$CDM}
\newcommand{\lcdm}{\LCDM\ }

\newcommand{\Rv}{R_{200}}
\newcommand{\RV}{$\Rv$}
\newcommand{\rv}{\RV\ }
\newcommand{\Mv}{M_{200}}
\newcommand{\MV}{$\Mv$}

\newcommand{\BM}{$\beta$-model}
\newcommand{\bm}{\BM\ }

\begin{document}

\maketitle

\begin{abstract}
Simple self-consistent models of galaxy groups and clusters are tested 
against the results of high-resolution adiabatic gasdynamical
simulations.
We investigate two models based on the existence of a 'universal' dark 
matter density profile and two versions of the beta-model.
The mass distribution of relaxed clusters can be fitted by
phenomenological formulae proposed in the literature.
Haloes that have experienced a recent merging event are
systematically less concentrated and show steeper profiles than
relaxed objects near the centre.
The hot X-ray emitting gas is found to be in approximate hydrostatic
equilibrium with the dark matter potential, and it is well described by
a polytropic equation of state.
Analytic formulae for the gas density and temperature can be derived
from these premises.
Though able to reproduce the X-ray surface brightness,
the beta-model is shown to provide a poor description
of our numerical clusters.
We find strong evidence of a 'universal' temperature profile that
decreases by a factor of $2-3$ from the centre to the virial radius.
We claim that the spherically-averaged profiles of all physical
properties of galaxy groups and clusters can be fitted with only two
free parameters.
Numerical resolution and entropy conservation play a key role in the
shapes of the simulated profiles at small radii.
\end{abstract}

\begin{keywords}
galaxies: clusters: general --- cosmology: theory --- methods: N-body
simulations
\end{keywords}

 \section{Introduction}
 \label{secIntro}

Clusters of galaxies are the largest gravitationally bound structures in the
universe, and as such they have often been considered as a canonical
data set for cosmological tests.
As a first-order approximation for analytical studies, they can be
described as spherically symmetric
systems in which the baryonic gas is in hydrostatic equilibrium with
an underlying cold dark matter (CDM) halo.

Even for this basic model, the radial distributions of both types of matter
are far from being well understood.
During the last two decades, a great effort has been devoted to
investigate the mass distribution of CDM haloes by means of numerical
N-body simulations. The systematic study undertaken by
\citet[hereafter NFW]{NFW97}
concluded that the simulated density profiles can be fitted by a
single two-parameter function, valid from galactic to cluster scales.

Similar results have been found in several independent papers,
although there is still some debate about the innermost value of the
logarithmic slope of the density profile and its dependence on resolution.
For instance, \citet{FukushigeMakino97,FukushigeMakino01},
\citet[][hereafter MQGSL]{Moore98,Moore99} or
\citet{Ghigna98_sh,Ghigna00_sh} find a
central slope steeper than the NFW profile.
On the other hand, \citet{JingSuto00} and \citet{Klypin01} claim that
the actual value might depend on halo mass, merging history and substructure.
More recently, \citet{Power03_sh} point out that the logarithmic slope
becomes increasingly shallower inwards, with little sign of
approaching an asymptotic value at the resolved radii.

The situation is even less clear for the baryonic component.
Most gas in the intracluster medium (ICM) is in the form of a hot 
diffuse X-ray emitting plasma, where the cooling time
(except in the innermost regions) is typically
longer than the age of the universe.
Adiabatic gasdynamical simulations have therefore been used to study
the formation and evolution of galaxy groups and clusters in different 
cosmologies \citep[e.g.][]{NFW95,EMN96,BN98,Eke98}.
This early work already showed that the gas distribution is
also similar in all objects, but more extended than that of the dark matter.
Temperature profiles are found to decrease significantly with radius,
but an isothermal core (or even increasing temperatures) of
variable extent have often been reported.

Recently, many numerical studies have focused on the effects
of radiative cooling and non-gravitational heating on the final properties
of groups and clusters of galaxies \citep[see
e.g.][]{Bialek01,Dave02,Muanwong02,Borgani01_sh,03T_sh}.
Entropy injection at high redshift and/or removal of
low-entropy gas modify the radial distribution of gas at small radii,
making the ICM even more extended than it would be
in the purely gravitational case.
Since cool systems are more sensitive to additional heating,
non-adiabatic processes have often been invoked to explain
the discrepancy between the observed $\Lx-\Tx$ relation
\citep[e.g.][]{EdgeStewart91} and the self-similar scaling
\citep{Kaiser86}, as well as the entropy excess detected in galaxy groups
\citep[e.g.][]{PCN99}.

The extent to which additional physics can influence the cluster
structure outside the cooling radius is unfortunately still unclear
\citep[see e.g.][]{Lewis00_sh,Pearce00,0302427}.
In this work, we are concerned with the theoretical predictions that
can be made about the radial structure of galaxy groups and clusters
beyond self-similarity arguments, in an
attempt to gain some understanding of the purely gravitational case.
We will consider that the ICM gas is in thermally-supported hydrostatic
equilibrium with the CDM halo, which in turn dominates the mass of the
system,
\be
\frac{1}{\rhogas(r)}\frac{\dd P(r)}{\dd r}=-\frac{GM(r)}{r^2}
\simeq-\frac{GM_{\rm dm}(r)}{r^2}
\label{eqHydro}
\ee
where $P=(\rhogas kT)/(\mu m_{\rm p})$ denotes gas pressure, $G$ is Newton's
constant, and $M$ is the mass enclosed within radius $r$.
$\rhogas$ and $T$ are the gas density and temperature,
$k$ is Boltzmann's constant, $m_{\rm p}$ is the proton mass and
$\mu$ is the mean molecular weight of an ionised plasma
of primordial composition ($\mu\simeq0.6$).

Several approximations are made in equation (\ref{eqHydro}). The
first one is spherical symmetry. Second, bulk and random motions
within the gas are not included. And third, baryons can contribute a
significant fraction of the total mass at large radii. These effects
can lead to appreciable departures from equation (\ref{eqHydro}) in
systems that are indeed in hydrostatic equilibrium. On the other hand, 
clusters of galaxies are evolving systems, and such a condition must
not necessarily be fulfilled, particularly by dynamically young objects.

Assuming a polytropic equation of state,
\be
P(r)\propto\rhogas^\gamma(r)
\label{eqPolyt}
\ee
a one-to-one correspondence between mass, gas density
and temperature can be established \citep{Makino98}.
Note that isothermality is a particular
case of a polytropic relation (i.e. $\gamma=1$).
The physical origin of a polytropic equation of state is still a
matter under investigation. A possible justification can be given in
terms of the entropy profile, related to the
mass accretion history of the system \citep{0304447}.

A different approach to the radial structure of galaxy groups and clusters
is the the so-called \BM,
pioneered by \citet{CavaliereFusco76} and widely used thereafter
\citep[see e.g.][for a recent review]{Rosati02}.
This model assumes that the ICM gas follows a 'universal' King-like
density profile
\be
\rhogas(r)=\rho_0\left[1+(r/r_{\rm c})^2\right]^{-3\beta/2}
\label{eqBetaRho}
\ee
and that its temperature is approximately constant.
Then, the observed X-ray surface brightness is given by the expression
\be
S_{\rm X}(\theta)=S_0\left[1+(\theta/r_{\rm c})^2\right]^{-3\beta+1/2}
\label{eqBetaS}
\ee
where $\theta$ is the projected radius from the observed X-ray
centroid and $S_0$ is the central surface brightness.

The physical meaning of $\beta$ is the ratio between gas
temperature and the line-of sight velocity dispersion of the CDM component.
The difference between this ratio ($\beta_{\rm spec}\sim1$) and the
fits obtained by applying (\ref{eqBetaS}) to observed clusters of
galaxies ($\beta_{\rm fit}\sim0.7$) has been often referred to as the
'$\beta$-discrepancy' \citep[see e.g.][]{Sarazin86}.
None the less, it has long been known both from
observations \citep[e.g.][]{NeumannArnaud99}
and simulations \citep[e.g.][]{BartelmannSteinmetz96}
that the \bm does not provide an optimal description of the gas
density, and that the best value of
$\beta_{\rm fit}$ tends to increase with the outermost radius used in
the fit.

In this paper, we investigate the radial distribution of dark and
baryonic matter by means of adiabatic gasdynamical simulations.
We assess the validity of the assumptions
of hydrostatic equilibrium and a polytropic equation of state,
and compare the density and temperature profiles
derived from these tenets with a sample of 15 numerical clusters of
galaxies.

The accuracy of the \bm has been previously addressed in several
numerical studies \citep[e.g.][]{EMN96,BartelmannSteinmetz96},
focusing mainly on the
mass estimates based on the observed X-ray surface brightness.
In the present work, we consider four analytical cluster models.
Two of them rely
on the existence of a 'universal' CDM density profile,
whereas the others assume a \bm for the gas distribution.
First, we attempt to obtain a self-consistent fit to our numerical data
with each one of these models, noting that baryonic and dark matter
profiles are not independent.
Then, we try to estimate several cluster profiles from the
simulated X-ray surface brightness.

We would like to make special emphasis on the
importance of numerical resolution and accurate entropy conservation.
We will show that the inner structure of the ICM depends critically on 
these issues. More specifically, we claim that they are responsible
for the systematic differences observed between
Eulerian and Lagrangian integration techniques \citep[see e.g.][]{SB_sh}.

Section~\ref{secSims} describes our numerical experiments.
Analytical cluster models are summarised in Section~\ref{secModels}.
Results are compared in Section~\ref{secResults}, and
the impact on the analysis and interpretation of current observations
is discussed in Section~\ref{secObs}.
We sum up our main conclusions in Section~\ref{secConclus}.

 \section{Numerical experiments}
 \label{secSims}

We have carried out a series of high-resolution gasdynamical
simulations of cluster formation in a flat CDM universe with non-vanishing
cosmological constant ($\Omega_{\rm m}=0.3$; $\Omega_\Lambda =0.7$;
$h=0.7$; $\sigma_8=0.9$; $\Omega_{\rm b}=0.02\ h^{-2}$). A thorough
description of the numerical experiments can be found in \citet{tesis}.

Simulations have been run with the parallel Tree-SPH code \g \citep{gadget01}.
We have used a new non-public version of the code, kindly provided by
Volker Springel, in which the entropy of SPH particles is explicitly
conserved \citep{gadgetEntro02}.
Similar experiments have been accomplished with the
{\sc Adaptive Refinement Tree} (ART) pure N-body code \citep{ART97}
from the same initial conditions.
In order to test the reliability of our results, one of the clusters
has been re-simulated with the gasdynamical version of ART
\citep{ARThydro02}.

 \subsection{Cluster sample}
 \label{secSample}

In a  cubic volume of 80 $h^{-1}$ Mpc on a side, an unconstrained
realization of the  power spectrum of density fluctuations
corresponding to the \lcdm model was generated for a
total of $1024^3$ Fourier modes. The  density field was then resampled
to a grid of $128^3$  particles, which were displaced from their
Lagrangian positions according to the Zeldovich approximation up to $z=49$.
Their evolution until the present epoch is traced by means of a pure
N-body simulation with $128^3$ dark matter particles.

Our sample comprises a total number of 15 clusters of galaxies,
selected from this preliminary low-resolution experiment.
Each object has been re-simulated with higher resolution by means of
the multiple mass technique \citep[see][for details]{Klypin01}. First,
we located the particles in a spherical region around the centre of
mass of the $128^3$ counterpart at $z=0$. Mass resolution is then
increased by using smaller masses in the Lagrangian volume
depicted  by these particles, including the additional small-scale
waves from the \lcdm power spectrum in the new initial conditions.

We use 3 levels of mass refinement, reaching an effective resolution
of $512^3$ CDM particles ($2.96\times 10^8\ h^{-1}$
M$_\odot$). Gas has been added in the highest resolved area
only. The total number of particles (dark+SPH) in this area is
greater than $2\times 10^6$ for all clusters.
The gravitational softening length was set to $\epsilon=2-5\ h^{-1}$ kpc, depending on number of particles within the virial radius \citep{Power03_sh}. The minimum smoothing length for SPH was fixed to the same value as $\epsilon$.
In total, 7 independent numerical experiments have been performed, running \g on a SGI Origin 3800 parallel computer at {\sc Ciemat} (Spain), using 32 CPU simultaneously. The average computing time needed to run  each  simulation was $\sim 8$ CPU  days ($6\times 10^5$ s). The same clusters have also been simulated with the N-body version of ART on the Hitachi SVR at the LRZ (Germany).

Table~\ref{tabSample} displays the physical properties of our
clusters at $z=0$. Objects have been sorted (and named)
according to their virial mass at the present day. Clusters J$_2$ and
K$_2$ are an exception to this rule, since they are two small groups
falling into J$_1$ and K$_1$, respectively.

\begin{table}
\begin{center}
\begin{tabular}{lccrrrr}
{\sc Cluster} & {\sc State} & \RV & \MV & $L_{200}^{\rm X}$
& $T_{200}$ & $T_{200}^{\rm X}$\\ \hline\\[-2mm]
~~   A    &  Minor  & 931 & 18.96 & 70.06 & 2.000 & 2.873 \\
~~   B    &  Minor  & 871 & 15.57 & 20.30 & 1.860 & 2.620 \\
~~   C    & Relaxed & 871 & 15.53 & 42.65 & 1.958 & 2.810 \\
~~   D    &  Minor  & 771 & 10.77 & 11.49 & 1.301 & 1.642 \\
~~   E    & Relaxed & 719 &  8.74 & 12.53 & 1.287 & 1.866 \\
~~   F    &  Major  & 661 &  6.79 & 12.47 & 1.059 & 1.195 \\
~~   G    &  Major  & 638 &  6.10 &  4.50 & 0.844 & 0.818 \\
~~   H    & Relaxed & 618 &  5.56 & 16.60 & 1.107 & 1.614 \\
~~   I    &  Major  & 581 &  4.60 &  2.34 & 0.666 & 0.666 \\
~~  J$_1$ & Relaxed & 584 &  4.67 &  9.99 & 0.937 & 1.498 \\
~~  K$_1$ & Relaxed & 557 &  4.06 &  4.98 & 0.690 & 1.043 \\
~~   L    &  Minor  & 547 &  3.84 &  1.21 & 0.624 & 0.779 \\
~~   M    &  Major  & 503 &  2.99 &  0.64 & 0.545 & 0.610 \\
~~ K$_2$  &  Minor  & 497 &  2.89 &  3.08 & 0.470 & 0.808 \\
~~ J$_2$  & Relaxed & 491 &  2.77 &  8.22 & 0.673 & 0.979 \\
\end{tabular}
\end{center}
\caption{Physical properties of the clusters at $z=0$, defined at
  200 times the critical density.
  Cluster name, dynamical state, \rv in $h^{-1}$ kpc,
  enclosed mass in $10^{13} h^{-1}$ \Msun, bolometric X-ray luminosity
  in $10^{25} h$ erg s$^{-1}$, and average temperatures (mass and
  emission-weighted) in keV.
  Individual images of our 15 clusters can be found in
  Appendix~\ref{secAtlas}.
}
\label{tabSample}
\end{table}

In order to study the effect of mergers and close encounters, we used
the {\sc Bound Density Maxima} galaxy finding algorithm \citep[see
e.g.][]{Klypin99,Colin99_sh}. We label as {\em major merger} any cluster in
which we are able to identify a companion structure inside \rv whose
mass is greater than 0.5 \MV;
if the most massive companion is in the range $[0.1-0.5\Mv]$,
the object is classified as a {\em minor merger}; otherwise, we
consider it a {\em relaxed} system in virial equilibrium.
The results of this classification scheme are quoted in the second
column of Table~\ref{tabSample}. Clusters named J$_i$ and K$_i$ are
relatively close pairs, but they are separate enough ($\sim2-3$ Mpc)
not to be considered as mergers.

We find a remarkable amount of dynamical activity in our randomly
selected sample. Only 6 objects have been catalogued as relaxed
clusters, whereas 5 fall into the category of minor mergers and 4 have
been classified as major merging systems. As pointed out by
\citet{Gottloeber01}, we find that the typical merging rate in groups
is much higher than in the more massive clusters.

Columns 3--7 display the bulk properties of each object.
The subscript '200' refers to the overdensity with respect to the
critical value, $\rho_c\simeq2.8\times10^{11}\ h^2$ \msun Mpc$^{-3}$.
Although the density contrast predicted in the spherical collapse
model is closer to 100 for our \lcdm cosmology, we chose this value
for consistency with most observational studies. Since most of the
mass is concentrated in the central regions, the difference between
the two prescriptions is not large.

Throughout this paper, we assume that X-ray emission arises 
from bremsstrahlung radiation only. The total luminosity of a set of
$N$ gas particles is therefore computed as
\be
\Lx=\lambda\sum_{i=1}^{N} m_i \rho_i T_i^{1/2}
\ee
where $\lambda=1.68\times10^{17}$ erg s$^{-1}$ M$_\odot^{-2}$ Mpc$^3$
keV$^{-1/2}$.
For equal mass particles, the emission-weighted temperature is given by
\be
\Tx=\frac{\sum_{i=1}^{N}\rho_iT_i^{3/2}}{\sum_{i=1}^{N}\rho_iT_i^{1/2}}
\label{eqTxN}
\ee
These formulae have been used to compute the corresponding entries in
Table~\ref{tabSample}, as well as the analytical and numerical
profiles of X-ray related quantities.

 \subsection{Comparison with other numerical techniques}
 \label{secSchemes}

Since the late 1980s a variety of techniques have been developed to
simulate gasdynamics in a cosmological context. In part inspired by
the success of the N-body scheme, the first gasdynamical techniques
were based on a particle representation of Lagrangian gas elements
using the smoothed particle hydrodynamics (SPH) technique
\citep{Lucy77,GingoldMonaghan77}.
Soon thereafter, fixed-mesh Eulerian methods were adapted
\citep{Cen90,Cen92} and, more recently, Eulerian methods with
sub-meshing \citep{BryanNorman95}, deformable moving meshes
\citep{Gnedin95,Pen95,Pen98} or adaptive mesh refinement
\citep[AMR,][]{Bryan95,ARThydro02} have been developed, as well as
extensions of the SPH technique \citep{Shapiro96,gadgetEntro02,deva}.

The Santa Barbara cluster comparison project \citep{SB_sh} attempted to
assess the extent to which existing modelling techniques gave
consistent results in a realistic astrophysical
application. The formation of a galaxy cluster in a SCDM universe was
simulated with 12 independent codes, seven of them based on the SPH
scheme and five on Eulerian methods.
The properties of the CDM component were encouragingly similar, most
discrepancies arising from small differences in timing (there was a
merging event at $z\sim0$). Less agreement was found in the
gas-related quantities, but usually still within $10-20$ per cent. Only the
X-ray luminosity showed a strong dependence on the resolution of the
different codes, with a spread as high as a factor of 10.

One of the most remarkable findings was a systematic trend in the
temperature profiles obtained for the inner regions ($r\leq100$ kpc). Near
the centre, SPH codes generate  a flat (or even slightly declining inwards)
temperature profile, while  grid codes produce temperature profiles
that are  still rising at the resolution limit.
 The entropy profile in  SPH codes decreases continously
towards the centre,  while grid codes develop 
an isentropic core at small radii.

Several authors
\citep[e.g.][]{Hernquist93,NelsonPapaloizou93,NelsonPapaloizou94,Serna96,deva,gadgetEntro02}
have pointed out that an important shortcoming of conventional SPH
formulations is the poor conservation of entropy when a low number of
particles is used. We have run several test simulations of the Santa
Barbara cluster to study the effects of numerical resolution and
entropy conservation. The standard
implementation of \g \citep{gadget01} gives similar results as the
SPH-based codes in \citet{SB_sh}, while the temperature and density
profiles of the entropy-conserving version \citep{gadgetEntro02} are
closer to the results of the Eulerian code by \citet{Bryan95} up to
the resolution limit \citep[see][for details]{tesis}. The innermost
radius $r_{\rm min}$ was defined according to the most restrictive of
the following criteria:
\begin{enumerate}
  \item 200 dark matter particles \citep{Klypin01}.
  \item 100 gas particles \citep{Borgani02_sh}.
  \item 3 times the gravitational softening \citep{Power03_sh}.
\end{enumerate}

Usually, the resolution limit is set by the last two conditions.
Excessively low values of the smoothing length can
lead to the formation of cold compact groups of 
SPH particles that decouple from the surrounding medium.
Although Eulerian simulations are not affected by this problem,
an unphysical temperature drop is often found in the cores of
clusters simulated with SPH-based codes at low and medium resolutions
\citep[e.g.][]{Eke98,SB_sh,ME01}. When the smoothing length is
properly set, a high number of particles is required to notice the
effects of numerical entropy loss, because the entropy profile
artificially flattens at $r\le r_{\rm min}$ due to the softened density
and temperature \citep[see e.g.][]{Borgani02_sh}.

\begin{figure}
  \centering \includegraphics[width=8cm]{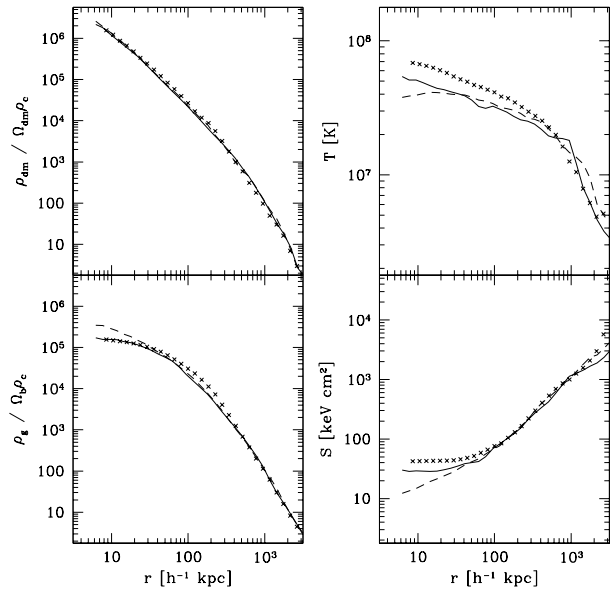}
  \caption{Simulations of Cluster A performed with the standard SPH
    (dashed lines) and entropy-conserving (solid lines)
    implementations of \G, compared to the results of the Eulerian
    code ART (crosses). \emph{Top left:} Dark matter density.
    \emph{Bottom left:} Gas density. \emph{Top right:} Gas
    temperature. \emph{Bottom right:} Gas entropy.
}
  \label{figA}
\end{figure}

The maximum resolution allowed by the Santa Barbara initial
conditions is only $256^3$ particles, and the
comparison between different runs is further complicated by the
merger at $z\sim0$.
Therefore, we decided to test the gasdynamical
integration scheme with the most massive of our clusters.
The results of the Eulerian code ART for Cluster A \citep{cluster6}
are compared with both implementations of \g in Figure~\ref{figA}.

As expected, CDM density profiles agree within a few
percent. Moreover, we find that the mass distribution in the pure N-body
ART runs is almost identical ($\Delta M/M\sim10$ per cent) to that
obtained in \G, once rescaled to account for the slightly different
values of $\Omega_{\rm dm}$ due to the presence of baryons. Only clusters
F, I, J and K display bigger discrepancies ($\sim40$ per cent) related to
offsets in timing (note that all these systems are major mergers).

Differences between both implementations of \g are mostly
evident in the gas distribution at small radii ($\sim20\ h^{-1}$ kpc).
The density profile is
steeper in the standard formulation of SPH, whereas the temperature is
systematically higher when entropy conservation is enforced. 
There is an excellent agreement between the gas density found in
ART and the entropy-conserving version of \G, but we find a
lower temperature in the latter throughout most of the cluster.

The shape of the temperature profile, though, is similar
in the two codes, and noticeably different from the approximately
isothermal structure found in the standard SPH version.
This affects the gas entropy near the centre, which shows
a 'floor' value of 30-40 keV cm$^{-2}$ at $r\sim20$ $h^{-1}$ kpc in
both ART and entropy-conserving \g runs, in contrast with most SPH
implementations (including the public version of \G), which tend to
produce a decreasing entropy profile down to the resolution
limit \citep[see e.g.][]{SB_sh}.

The agreement between the Eulerian code ART and the explicit
entropy-conserving scheme by \citet{gadgetEntro02} provides
encouraging support to this new SPH formulation.
Furthermore, it constitutes a sound piece of evidence that the
conventional scheme
suffers from severe entropy losses, with important consequences
on the shape of the profiles in the
central regions of groups and clusters of galaxies.
We conclude that a careful modelisation of adiabatic processes is an
essential requisite prior to the inclusion of additional physics, such 
as cooling or preheating of the intergalactic medium at high redshift.

 \section{Analytical models}
 \label{secModels}

We compare our numerical data with self-consistent analytical cluster
models based on the assumptions of hydrostatic equilibrium and a
polytropic equation of state.
As a first step, we check whether these
approximations are consistent with our simulated sample of galaxy
groups and clusters.

Once the reliability of the basic assumptions is tested, we focus on
two different kinds of models: one is based on the existence of a
'universal' dark matter density profile (e.g. NFW or MQGSL) and the
other relies on a 'universal' gas density (i.e. \BM).
In total, we test four different descriptions of the radial structure
of galaxy groups and clusters. Hereafter, we will refer to them as
NFW, MQGSL, \bm (BM), and polytropic \bm (PBM). 

 \subsection{Basic assumptions}

\begin{figure}
  \centering \includegraphics[width=8cm]{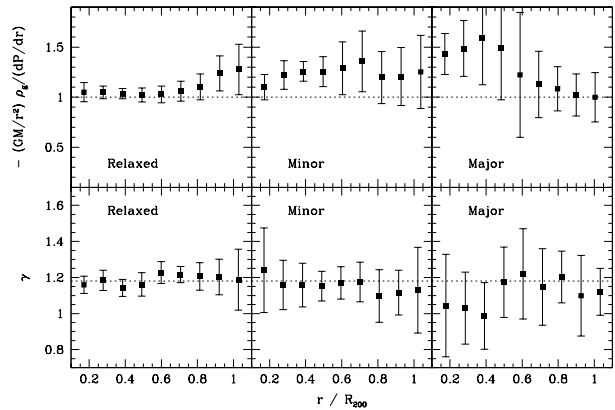}
  \caption{Hydrostatic equilibrium and polytropic equation of state in
    our numerical sample.
    Objects have been classified according to their dynamical state 
    as explained in~\ref{secSample}.
    Solid squares correspond to the average over all systems in each
    category, and error bars are used to indicate one-sigma deviation of
    individual profiles (dotted lines) around the average.
    \emph{Top panel:} Test of the hydrostatic equilibrium assumption.
    \emph{Bottom panel:} Polytropic index as a function of $r$.
}
  \label{figAssum}
\end{figure}

The extent to which equations (\ref{eqHydro}) and (\ref{eqPolyt})
hold for our numerical clusters is measured in Figure~\ref{figAssum}.
On the top panel, we plot the ratio between the gravitational term
$GM/r^2$ and the pressure gradient  $\rhogas^{-1}\dd P/\dd r$.
For a gas in thermally-supported hydrostatic equilibrium
(i.e. neglecting infall, angular momentum and turbulence), both
quantities should be equal to satisfy equation (\ref{eqHydro}).
Clusters that have been classified as relaxed on dynamical grounds can
be considered in hydrostatic equilibrium with an accuracy better than
$10$ per cent. For minor mergers, this assumption holds only
marginally ($\sim25$ per cent).
Clusters undergoing a major merger display an even larger scatter,
particularly at small radii.

Deviations from hydrostatic equilibrium arise mostly from the
contribution of  kinetic energy to the gravitational support of the
ICM gas. In relaxed clusters, thermal energy dominates the total
energy budget.
Velocity dispersion increases substantially during merging events, and
the gas needs some time to dissipate this extra kinetic energy into heat.
Since relaxed systems constitute only $40$ per cent of our sample, some
caution must be kept in mind whenever equation (\ref{eqHydro}) is
applied to a real cluster ensemble. We note as well that infall and
random motions are important in relaxed clusters for $r\ge0.8\Rv$.

Concerning the equation of state, the most straightforward way to
check whether the ICM follows a polytropic relation is to compute the
effective value of the polytropic index from the spherically-averaged
density and temperature profiles,
\be
\gamma(r)=1+\frac{\dd\log[T(r)]}{\dd\log[\rhogas(r)]}
\ee

\begin{figure}
  \centering \includegraphics[width=8cm]{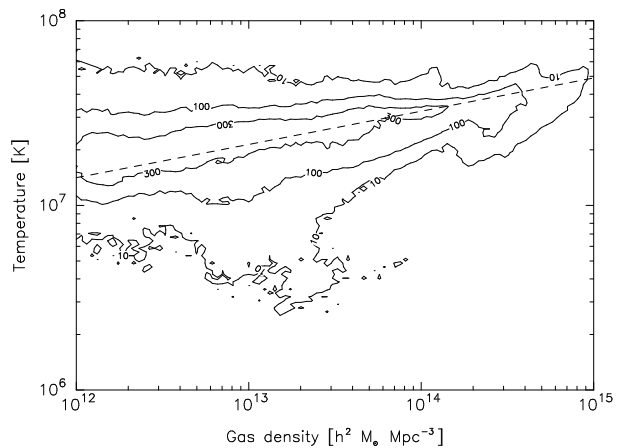}
  \caption{Histogram of individual gas particles of Cluster A in
    the $\rhogas-T$ plane. Contour lines are drawn at 10, 100 and 300
    particles per bin. Dashed line shows the expected
    slope for $\gamma=1.18$.}
  \label{figPolytA}
\end{figure}

This quantity is shown on the bottom panel of
Figure~\ref{figAssum}. Albeit some scatter, gas in our relaxed
clusters and minor mergers is consistent with a polytropic equation of
state with $\gamma\sim1.18$ (dashed line). An isothermal profile
($\gamma=1$) can be confidently ruled out from these data.
A density-temperature histogram of the gas particles of Cluster A is
plotted in Figure~\ref{figPolytA}.
We see that a polytropic equation
of state reflects not only the mean behaviour of the ICM gas, but also 
describes the actual ratio between the density and temperature of
individual mass elements.
Although only one of our clusters is shown in Figure~\ref{figPolytA},
the plot looks similar for any other object.

Gas in small haloes is typically much colder than the intracluster
medium.
Cold gas clumps corresponding to infalling galaxies can
be distinctly appreciated in Figure~\ref{figPolytA}, but their
mass is too low to alter the effective polytropic index 
of Cluster A.
However, the profiles become very noisy beyond the virial radius,
as well as in major mergers. A constant $\gamma$ is not an accurate
approximation in these cases, where the amount of substructure within
a spherical shell is large.

 \subsection{Cluster models based on a 'universal' CDM density profile}

As pointed out in \citet{Makino98} and \citet{Suto98}, the gas density
and temperature profiles can be derived analytically from the CDM mass
distribution under the assumptions of hydrostatic equilibrium and a
polytropic equation of state.
In the present work, we will consider the functional forms proposed by NFW,
\be
\rhodm^{\rm NFW}(r)=\frac{\rhos}{(r/\rs)(1+r/\rs)^2}
\label{eqNFW}
\ee
and MQGSL,
\be
\rhodm^{\rm MQGSL}(r)=\frac{\rho_{\rm m}}{(r/r_{\rm m})^{3/2}\left[1+(r/r_{\rm m})^{3/2}\right]}
\label{eqMQGSL}
\ee

Gas density and temperature profiles can be computed by substituting
these mass distributions into equation (\ref{eqHydro}) and combining
it with equation (\ref{eqPolyt}). Following the notation of
\citet{Suto98}, the gas temperature is given by
\be
\frac{T(r)}{T_0}=1-B_{\rm i} f_{\rm i}(r/r_{\rm i})
\ee
where the subscript 'i' can be either 's' or 'm' for NFW and MQGSL models,
\be
f_{\rm s}(x)=1-\frac{\ln(1+x)}{x}
\ee
and\footnote{Note the typos in formulas (15) and (16) of \citet{Suto98}.}
\bea
\nonumber
  f_{\rm m}(x)&=& \frac{\pi}{3\sqrt{3}}
  + \frac{2}{\sqrt{3}}\arctan\left(\frac{2x^{1/2}-1}{\sqrt{3}}\right)\\
& &
  + \frac{1}{3}\ln\left(\frac{x+2x^{1/2}+1}{x-x^{1/2}+1}\right)
  - \frac{2\ln(1+x^{3/2})}{3x}
\label{eqFMQGSL}
\eea
In both cases,
\be
B_i=
\frac{4\pi G\mu m_{\rm p}(\gamma-1)\rho_{\rm i} r_{\rm i}^2}
{\gamma kT_0}
\label{eqBi}
\ee
relates the central temperature $T_0$ to the underlying dark matter
distribution. The gas density can be easily obtained from the 
polytropic relation
\be
\frac{\rhogas(r)}{\rho_0}=\left[\frac{T(r)}{T_0}\right]^{1/(\gamma-1)}
\ee

According to this simple picture, clusters of galaxies can be
described as a function of 5 parameters. A characteristic
density and radius define the properties of the CDM halo,
whereas for the ICM we also need to specify the gas density and
temperature at $r=0$, as well as the polytropic index $\gamma$.

However, equation (\ref{eqBi}) relates the values of these variables.
The number of free parameters can be further reduced by imposing additional
constraints. For instance, \citet{KS01} choose $T_0$ and $\gamma$ by
enforcing approximately constant baryon fraction at large radii.
In our case, we just impose vanishing density at
infinity, i.e. $B_{\rm s}=1$ and $B_{\rm m}=3\sqrt{3}/(4\pi)$, and we
compute the central gas density from the condition that the baryon
fraction never exceeds the cosmic value,
\be
\frac{\rhogas(r_{\rm max})}{\rhodm(r_{\rm max})}
=\frac{\Omega_{\rm b}}{\Omega_{\rm dm}}
\ee
where $r_{\rm max}$ is the radius at which the predicted baryon
fraction reaches its maximum ($2.71\rs$ or $1.34r_{\rm m}$).
With this prescription,
\be
\rho_0^{\rm NFW}  \simeq
        1.51\frac{\Omega_{\rm b}}{\Omega_{\rm dm}}\rhos ~~,~~
\rho_0^{\rm MQGSL}\simeq
        23.08\frac{\Omega_{\rm b}}{\Omega_{\rm dm}}\rho_{\rm m}
\label{eqFbNFW}
\ee

In principle, there is no reason why the baryon fraction cannot exceed
the cosmic value somewhere within the cluster.
Yet, our numerical results (see Section~\ref{secGasDens} below) indicate
that the gas to dark matter ratio increases monotonically
(i.e. $r_{\rm max}=\infty$) up to $\Omega_{\rm b}/\Omega_{\rm dm}$.
Since the analytical estimate obtained for NFW and MQGSL models does not
tend to an asymptotic value, we decided to set the normalisation at the
maximum. The value of $\rho_0$ varies by $\sim 25$
per cent for any choice between $r_{\rm max}$ and \RV, but
inaccuracies in the assumption of thermally-supported hydrostatic
equilibrium might bias a prescription based on large radii.

The last parameter in this family
of models is the polytropic index $\gamma$.
According to our results (see Figure~\ref{figAssum}),
we set it to $\gamma=1.18$. Although this value is consistent with
recent observations \citep[e.g.][]{Markevitch98,Sanderson03_sh}, it
would be interesting
to investigate its physical origin, as well as whether it can be
related to other parameters such as
halo mass or concentration. However, such a study requires a
much larger number of objects, over a broader mass range, in order to
be statistically significant.

 \subsection{$\beta$-models}

Most observational studies do not explicitly rely on the assumption of
a given CDM density profile. Instead, they fit the observed X-ray
surface brightness with equation (\ref{eqBetaS}), and then compute the
dark and baryonic profiles by applying hydrostatic equilibrium and a
polytropic (often isothermal) equation of state.

To a great extent, the so-called '$\beta$-discrepancy' is due to the
fact that expression (\ref{eqBetaRho}) does not provide a good fit to
the gas distribution throughout the whole cluster \citep[see
e.g.][]{NFW95,BartelmannSteinmetz96}.
\citet{ME01} argue that line emission from cold gas in the
outskirts of the cluster is entirely responsible for the observed
value $\beta\sim2/3$.

Polytropic $\beta$-models have also been proposed to account for the
presence of a temperature gradient, but equation (\ref{eqBetaRho}) is
still used to model the gas density \citep[e.g.][]{Ettori00}.
In the general polytropic case, the surface brightness
\be
S_{\rm X}(\theta)=S_0\left[1+(\theta/r_{\rm c})^2\right]^
{1/2-3\beta(1+\frac{\gamma-1}{4})}
\label{eqBetaSpolyt}
\ee
is similar to the conventional form (\ref{eqBetaS}) for any reasonable 
value of the polytropic index ($1<\gamma<5/3$). Therfore, the gas
density inferred from a fit to the observed X-ray profile has virtually 
the same shape in both the isothermal (BM) and polytropic (PBM) \BM s.
The central surface brightness is given by
\be
S_0=\sqrt{\pi}\rho_0^2r_{\rm c}\lambda T_0^{1/2}
    \frac{\Gamma[3\beta(1+\frac{\gamma-1}{4})-1/2]}
         {\Gamma[3\beta(1+\frac{\gamma-1}{4})]}
\label{eqS0}
\ee

Hydrostatic equilibrium leads to an underlying mass distribution
\be
M(r)=\frac{3\beta kT_0\gamma r_{\rm c}}{G\mu m_{\rm p}}
(r/r_{\rm c})^3\left[1+(r/r_{\rm c})^2\right]^{-3\beta(\gamma-1)/2-1}
\ee
which features a finite central density
\be
\rhodm(0)\equiv\rho_\beta=
\frac{9\beta kT_0\gamma}{4\pi G\mu m_{\rm p}r_{\rm c}^2}
\label{eqRhoBeta}
\ee

Therefore, the structure of the dark matter halo in the \bm is related to
gas temperature in a way analogous to the models described above.
We can follow the same procedure to normalise the central gas density,
imposing that the maximum baryon fraction equals the cosmic value.
We will consider a standard \bm ($\gamma=1$) with $\beta=2/3$ and a
polytropic form with $\gamma=1.18$ and $\beta=1$. For these values,
the maximum baryon fraction occurs at $r_{\rm max}^{\rm BM}\to\infty$
and $r_{\rm max}^{\rm PBM}\simeq4.18r_{\rm c}$, respectively, and hence
\be
\rho_0^{\rm BM}=\frac{1}{3}
        \frac{\Omega_{\rm b}}{\Omega_{\rm dm}}\rho_\beta ~~,~~
\rho_0^{\rm PBM}\simeq0.39
        \frac{\Omega_{\rm b}}{\Omega_{\rm dm}}\rho_\beta
\label{eqFbBM}
\ee

\begin{table*}
\begin{center}
\begin{tabular}{cccc}
\makebox[2cm]{} & NFW & MQGSL & BM / PBM \\ \hline\\[-2mm]
$\rhodm(x)$
         & $x^{-1}(1+x)^{-2}$
         & $x^{-3/2}(1+x^{3/2})^{-1}$
         & $\left[1+(\frac{1}{3}-B)x^2\right](1+x^2)^{-3B/2-2}$ \\[2mm]
$M_{\rm dm}(x)$
         & $\ln(1+x)-x(1+x)^{-1}$
         & $\frac{2}{3}\ln(1+x^{3/2})$
         & $\frac{1}{3}x^3(1+x^2)^{-3B/2-1}$ \\[2mm]
$\rhogas(x)$
         & $\left[x^{-1}\ln(1+x)\right]^{1/(\gamma-1)}$
         & $\left[1-\frac{3\sqrt{3}}{4\pi}f_{\rm m}(x)\right]^{1/(\gamma-1)}$
         & $(1+x^2)^{-3\beta/2}$\\[2mm]
$T(x)$
         & $x^{-1}\ln(1+x)$
         & $1-\frac{3\sqrt{3}}{4\pi}f_{\rm m}(x)$
         & $(1+x^2)^{-3B/2}$
\end{tabular}
\end{center}
\caption{Polytropic models of galaxy clusters in hydrostatic
  equilibrium. $x$ denotes the radial coordinate in 
  units of the characteristic radius of each model ($\rs$ in NFW,
  $r_{\rm m}$ in MQGSL, and $r_{\rm c}$ in \BM s).
  Dark matter density is scaled in terms 
  of the characteristic density ($\rhos$, $\rho_{\rm m}$, and
  $\rho_\beta$, respectively).
  Masses are in units of $4\pi\rho_{\rm i}r_{\rm i}^3$.
  Gas density and temperature are scaled by
  their central values $\rho_0$ and $T_0$.
  The abbreviations $f_{\rm m}(x)$, given by (\ref{eqFMQGSL}),
  and $B\equiv\beta(\gamma-1)$ have been used.
}
\label{tabModels}
\end{table*}

We quote in Table~\ref{tabModels} the distributions of dark and
baryonic matter expected in our four analytical cluster models.
Cumulative gas mass and projected quantities in
NFW and MQGSL must be integrated numerically.
Relations between the characteristic parameters 
are summarised in Table~\ref{tabModels2}.

\begin{table}
\begin{center}
\begin{tabular}{cccc}
{\sc Model} & $S_0/(\lambda\rho_0^2T_0^{1/2}\ri)$
            & $(\Omega_{\rm b}\rhoi)/(\Omega_{\rm dm}\rho_0)$
            & $T_0/(4\pi G\mu m_{\rm p}\rhoi\ri^2)$
\\ \hline\\[-2mm]
NFW      & 0.191  &  1.51 & 0.153\\
MQGSL    & 0.0151 & 23.08 & 0.369\\
BM       & 1.571  &  1/3  & 6 \\
PBM      & 1.145  &  0.39 & 10.62 \\
\end{tabular}
\end{center}
\caption{Relations between the characteristic radius $\ri$, CDM
  density $\rhoi$, gas density $\rho_0$, and gas temperature $T_0$.
  All quantities have been computed assuming $\gamma=1.18$, except in
  the BM ($\gamma=1$). For this model, $\beta=2/3$; for PBM, $\beta=1$.
}
\label{tabModels2}
\end{table}

 \section{Results}
 \label{secResults}

We have shown that our clusters are, to a fair extent, in
hydrostatic equilibrium, and that a polytropic equation of state with
$\gamma=1.18$ provides a good description of the hot diffuse component 
of the ICM.
Under these conditions, models of the radial structure of galaxy
groups and clusters can be derived from phenomenological approaches to 
either the gas or dark matter density.
In this Section, we compare these simple, self-consistent models with
the spherically-averaged distributions of mass, gas density
and temperature found in our sample of simulated galaxy clusters.

For NFW and MQGSL, we fit the cumulative dark mass of each halo,
while for the \BM s the gas mass is used instead.
We compute the numerical profiles in 26 logarithmic bins
between $0.05\Rv$ and \RV.
We set the lower cut-off well above our resolution limit
($\sim0.01\Rv$) not only to
avoid numerical effects but also to investigate how accurate are the
extrapolations of the analytical cluster models towards $r=0$.

The parameter space is explored by
letting $r_{200}$ and $c_{\rm i}\equiv r_{200}/r_{\rm i}$ vary
uniformly in the range $0.25<r_{200}<1.25\ h^{-1}$ Mpc
and $0.1<c_{\rm i}<16$ (50 for BM).
For the gas mass, the radius $r_{10}$ (enclosing a mean
density of $10\rho_{\rm c}$) is used instead of $r_{200}$. Both quantities are roughly
equivalent for the baryon fraction assumed in this work.

\begin{figure}
  \centering
\includegraphics[width=8cm]{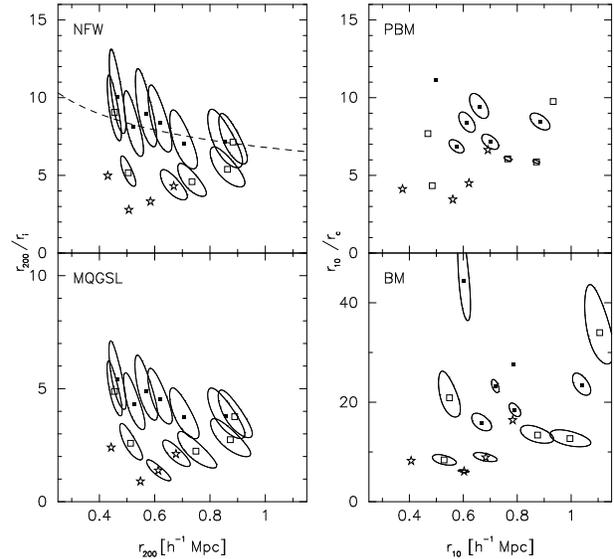}
  \caption{Best-fitting model parameters for our sample of numerical
    galaxy clusters (see text for details on the fitting procedure).
    Solid squares represent relaxed objects, empty squares are used
    for minor mergers and stars for major merging systems.
    Contours indicate
    $\sqrt{\chi^2/(dof)}=0.1$.
    Dashed line in NFW panel plots $c=6.8r_{\rm 200}^{-0.3}$.
}
  \label{figFit1}
\end{figure}

Results of the $\chi^2$ minimisation are plotted
in Figure~\ref{figFit1}.
In most cases, the quality of the fit is only slightly better
for NFW and MQGSL models ($\sqrt{\chi^2/(dof)}\sim0.05-0.1$)
than for \BM s ($0.05-0.15$).
Best-fitting values of $r_{200}$ are close to the actual \rv
(quoted in Table~\ref{tabSample}) obtained directly from the simulated
mass profile.
The values of $r_{10}$ in BM ($\beta=2/3$) are less
well correlated with \rv than those obtained for PBM ($\beta=1$).

Although the full sample is consistent with constant dark matter
concentration over the restricted mass range probed,
we notice a well-defined bimodal behaviour:
while relaxed clusters follow the usual relation,
approximated by $c=186\Mv^{-0.1}$ for the NFW profile
\citep{BurkertSilk99_sh}, mergers are dramatically offset
down, particularly low-mass systems.
In the \BM s, we find that the best-fitting gas concentration even 
\emph{increases} with cluster mass. In several systems, the core
radius inferred by the BM is only slightly larger than the innermost
radial bin included in the fit.

An important fact is that dark and baryonic matter distributions are
expected to be strongly correlated, by virtue of the hydrostatic
equilibrium equation and the constraint of a cosmic baryon fraction.
All analytical models of galaxy groups and
clusters studied in the present work have only two free parameters.
According to our fitting procedure,
gas density and temperature are genuine predictions (i.e. not fits)
of the models based on a dark matter profile.
Conversely, the same is true for the CDM density and gas temperature
in the \BM s.

 \subsection{Mass distribution}
 \label{secDM}

\begin{figure}
  \centering \includegraphics[width=8cm]{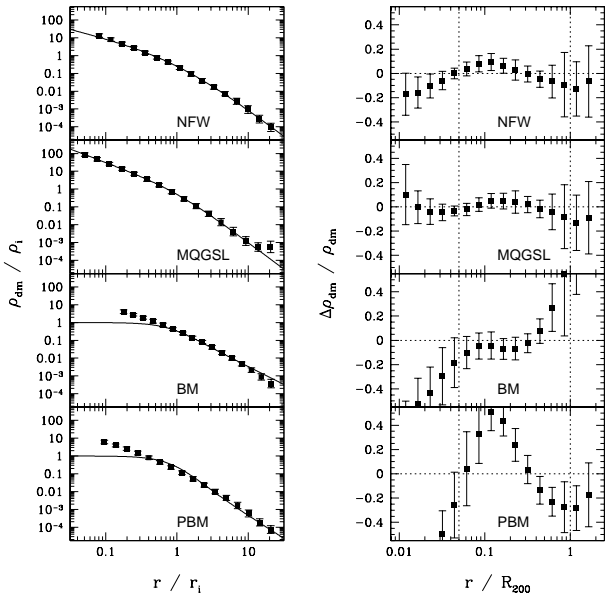}
  \caption{Dark matter density profile of our numerical haloes.
    Black squares represent the average over all clusters, excluding
    major mergers. Error bars indicate one-sigma scatter of individual 
    profiles.
    \emph{Left panel:} CDM density, scaled by the
    best-fitting $\rhoi$ and $\ri$ of each object in the
    corresponding model.
    Analytical profiles are shown as solid lines.
    \emph{Right panel:} Accuracy of each analytical
    profile. Vertical dotted lines mark the fitted region.
}
  \label{figDM1}
\end{figure}

The radial density profiles of our dark matter haloes
are shown on the left panel of Figure~\ref{figDM1},
rescaled by their best-fitting characteristic densities and radii.
We plot the expected mass distribution in each model
(see Table~\ref{tabModels}) as a solid line.
On the right panel, we plot the accuracy of the analytical estimates,
defined as
\be
\frac{\Delta\rhodm}{\rhodm}\equiv
\frac{\rho_{\rm est}(r)-\rho_{\rm num}(r)}{\rho_{\rm num}(r)}
\ee
where $\rho_{\rm est}$ and $\rho_{\rm num}$ are the estimated and
numerical profiles, respectively.

Not surprisingly, both NFW and MQGSL offer an excellent fit to the
dark matter distribution, but the core-like predictions of the \BM s
represent a very poor approximation to the numerical profiles.
The conventional \bm has a much better accuracy throughout the fitted
range, but the extrapolation to larger radii is more reliable in the
polytropic version, thanks to its higher value of $\beta$.
\footnote{In the isothermal case, though, dark matter density is
  insensitive to the actual value of this parameter (see
  Table~\ref{tabModels}).}

\begin{figure}
  \centering \includegraphics[width=8cm]{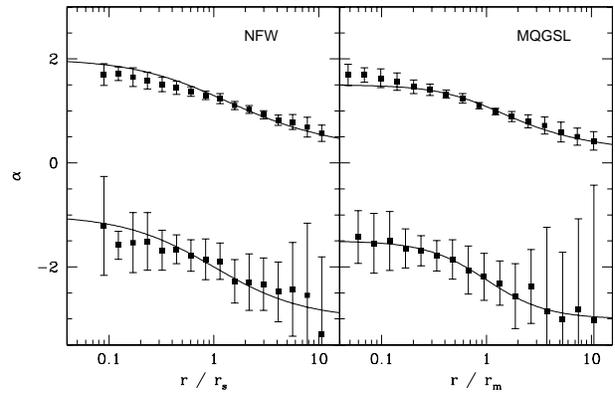}
  \caption{Logarithmic slope of the CDM mass (top) and density
    (bottom) profiles.
    Squares with error bars represent the average over relaxed
    clusters and minor mergers.
    Analytical cluster models are shown as lines.
    }
  \label{figDM2}
\end{figure}

Concerning the controversy about the inner slope of the density profile,
it is apparent from Figure~\ref{figDM1} that MQGSL formula
is slightly more accurate than NFW near the centre, albeit the scatter 
of individual profiles around the average is somewhat higher.
This issue is addressed in more detail in Figure~\ref{figDM2}, where
we plot the logarithmic slopes of the CDM density and cumulative mass
profiles, computed directly from the simulation data
\be
\alpha_{\rho} = \frac{\dd\log[  \rhodm(r)  ]}{\dd\log(r)} ~~,~~
\alpha_{\rm M}= \frac{\dd\log[M_{\rm dm}(r)]}{\dd\log(r)}
\ee

In general terms, both NFW and MQGSL formulae are consistent with our
results up to the resolution limit.
When we divide our sample according to the dynamical
state of each cluster, we find a well defined trend in the
sense that relaxed haloes display shallower central slopes (close to
NFW) than merging systems (better described by MQGSL).
As an extreme case, major mergers feature pure
power-law mass distributions for more than one decade in radius, whose 
slopes are even steeper than the value proposed by MQGSL.
In agreement with \citet{Power03_sh}, we do not find any sign of an
asymptotic slope for either relaxed clusters or minor mergers.
More resolution is clearly required before achieving a firm
conclusion on this matter.


\subsection{Gas temperature}

Mass determinations from X-ray emission in clusters usually assume
that relaxed clusters (i.e. morphologically symmetric) are
isothermal. However, this assumption has been questioned recently by
\citet{Markevitch98}, who found evidence of a decreasing temperature
profile for nearby clusters observed with ASCA. This result has been
confirmed for cool clusters by \citet{Finoguenov01}. On the contrary,
\citet{White00} and \citet{IrwinBregman00}, using data from {\em
  Beppo}SAX and ASCA satellites, do not find any decrease of the
temperature in a large collection of clusters. More recently, the
analysis of {\em Beppo}SAX observations accomplished by
\citet{GrandiMolendi02} concluded that temperature profiles of galaxy
clusters can be described by an isothermal core followed by a rapid
decrease.

Numerical simulations of galaxy clusters support a significant
decrease in the ICM temperature from the central regions to the virial
radius \citep[e.g.][]{SB_sh}.
Recent results from high resolution AMR gasdynamical simulations seem
to indicate that the temperature profile has an universal form
\citep{Loken02_sh}. These authors proposed a simple formula to fit the
projected emission-weighted temperature of their clusters:
\be
 T_{\rm p}^{\rm X}(\theta)=T_0{(1+\theta/\ax)}^{-\delta}
\label{eqLoken}
\ee
where $T_0$ is the central temperature and $\ax$ a core
radius. \citet{Loken02_sh} quote best-fitting values of these parameters
$\ax=r_{\rm vir}/1.5$ and $\delta=1.6$.

\begin{figure}
  \centering \includegraphics[width=8cm]{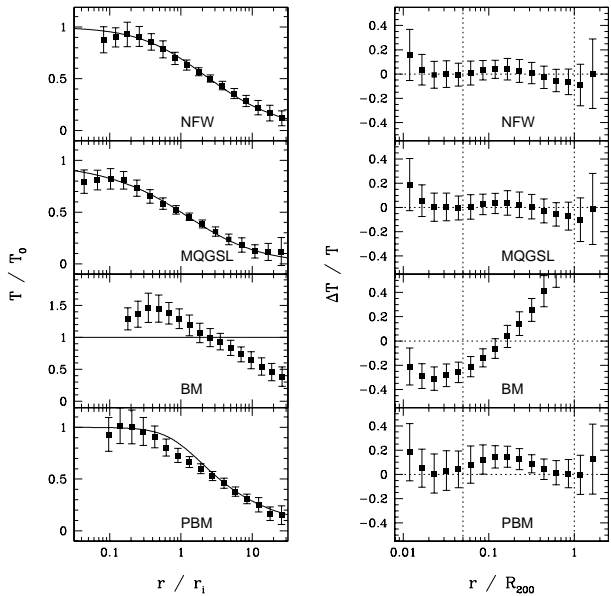}
  \caption{Same as Figure~\ref{figDM1}, for the temperature profile.}
  \label{figT}
\end{figure}

The average mass-weighted temperature profile of our sample of galaxy
clusters (excluding major mergers) is shown on the left panel of  Figure~\ref{figT}.
The normalisation $T_0$ has been computed analytically from the fit to
the dark matter (NFW, MQGSL) or gas (BM, PBM) distribution of each
cluster.
As can be seen on the right panel, the theoretical prediction is
accurate within $\sim15$ per cent for
all prescriptions, with the obvious exception of the isothermal \BM.
NFW and MQGSL models are virtually indistinguishable.

\begin{figure}
  \centering \includegraphics[width=8cm]{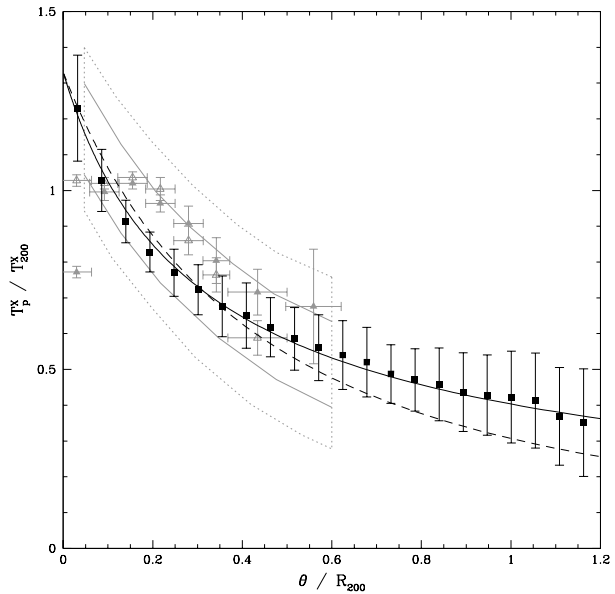}
  \caption{Projected emission-weighted temperature profile, scaled by
    $T_{200}^{\rm X}$ in Table~\ref{tabSample} (black squares with
    error bars). Solid line is the best fit obtained with equation
    (\ref{eqLoken}), dashed line represents the simulation results of
    \citet{Loken02_sh}. Observational data from \citet{GrandiMolendi02}
    are shown as triangles (filled for cooling flow and empty for
    non-cooling flow clusters). Dotted box encloses the temperature
    profiles observed by \citet{Markevitch98} plus most of the error bars,
    solid box includes only the scatter of their best-fitting
    polytropic models.}
  \label{figTx}
\end{figure}

In order to compare with observations, as well as with the numerical work
of \citet{Loken02_sh}, we plot in Figure~\ref{figTx} the average
projected X-ray temperature profile.
We computed the emission-weighted temperature according to equation
(\ref{eqTxN}) in cylindrical shells of 3 $h^{-1}$ Mpc length, oriented
along the main axes of the simulation box.
The final profile of each cluster is given by the average over the
three orthogonal projections, and the normalisation comes from the
emission-weighted temperature, $T_{200}^X$,
within \rv (both quantities are quoted in Table~\ref{tabSample}).

Our results are in excellent agreement with the observational
estimate of \citet{Markevitch98}, who claim that the
emission-weighted temperature is well
represented by a polytropic $\beta$-model (indeed, our data are almost
identical to their best-fitting profile).
However, our profiles are only marginally consistent
with the results reported by \citet{GrandiMolendi02}.
In particular, we do not find any evidence of the temperature
flattening observed by these authors at $r\sim0.2\Rv$.

The projected X-ray temperature of our clusters rises uninterruptedly
until the innermost radial bin, in agreement with the results of
Eulerian codes, such as ART (see Section~\ref{secSchemes})
or that employed by \citet{Loken02_sh}.
Equation (\ref{eqLoken}) fits well our numerical data, but we obtain
rather different values of the core radius and the asymptotic
exponent, $\ax=r_{200}/4.5$ and $\delta=0.7$.
Since both simulated temperature profiles are consistent
within the error bars, it seems that the best-fitting values of
$\ax$ and $\delta$ are rather sensitive to the details of the data.

Never the less, we would like to stress that a decreasing temperature
profile is a robust result of numerical simulations regardless of the
integration scheme (either AMR or SPH), as long as entropy
conservation is enforced in the latter.
Furthermore, such behaviour is strongly supported
by most recent observational measurements.
The presence of an isothermal core in the observed galaxy clusters could
reflect the need for additional non-adiabatic physics in the
simulations.

 \subsection{Gas density}
 \label{secGasDens}

\begin{figure}
  \centering \includegraphics[width=8cm]{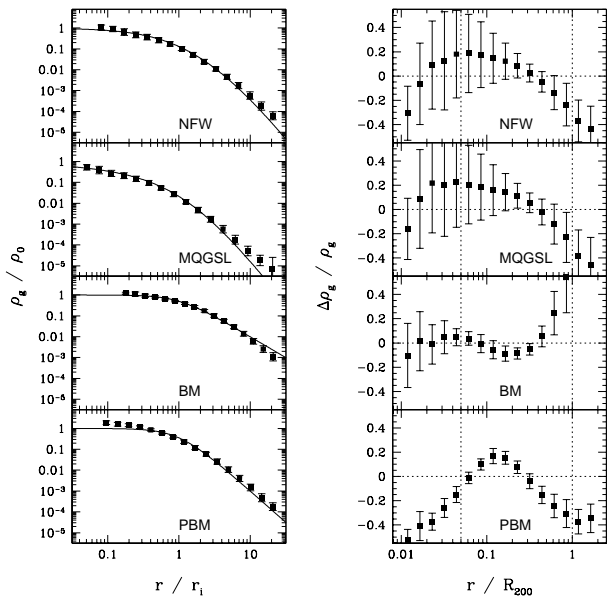}
  \caption{Same as Figure~\ref{figDM1}, for the gas density profile.}
  \label{figD}
\end{figure}

Finally, we compare the simulated gas density profile with the
distributions expected in our four analytical cluster models.
It is evident from Figure~\ref{figD} that the discrepancies are
remarkably larger than for the dark matter density or ICM temperature
profiles.

It is somewhat surprising that the \BM s show the worst
inconsistencies with the data, since they are specifically intended 
to fit the gas distribution.
When $\beta=2/3$, the asymptotic gas density falls
too slowly, as $r^{-2}$. However, the central regions cannot be
properly fitted if $\beta=1$ is assumed.
The final outcome is that the best-fitting value of $\beta$ increases
from one value to the other as larger radii are considered in the fit.

Models based on a 'universal' dark matter profile provide a fair
description of the ICM density.
None the less, we caution that in many cases the simulated gas density
can differ substantially from the analytical prediction, giving rise
to the huge scatter ($\sim40$ per cent) observed in Figure~\ref{figD}.
On one hand, the CDM distribution at small radii depends on the
dynamical state of the halo.
Near the virial radius, even relaxed
clusters deviate noticeably from thermally-supported hydrostatic
equilibrium (see Figure~\ref{figAssum}), and
we also observe that the dark matter density is slightly higher than
indicated by both NFW and MQGSL formulae (see Figure~\ref{figDM1}).
More importantly, the polytropic index is not exactly the same for all 
clusters, which strongly alters the shape of the inferred gas
density profile.

\begin{figure}
  \centering \includegraphics[width=8cm]{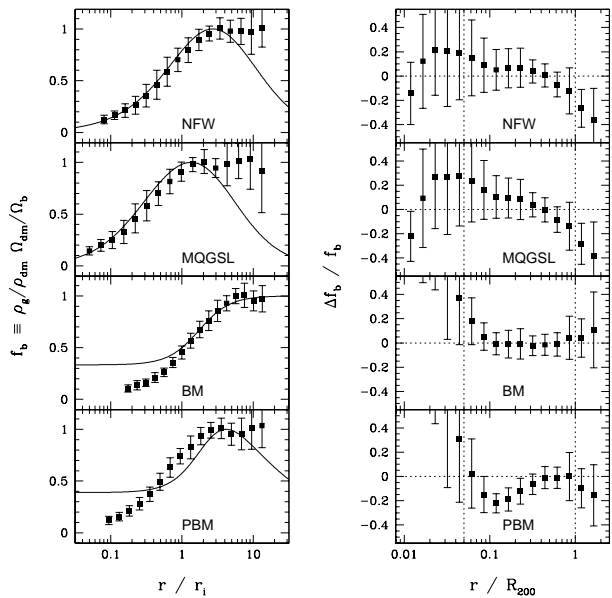}
  \caption{Same as Figure~\ref{figDM1}, for the local baryon fraction.}
  \label{figfb}
\end{figure}

Inaccuracies in the estimated dark matter and/or gas profiles lead to
a significant mismatch between the analytical and numerical baryon
fractions.
The local ratio of gas to CDM density is plotted in Figure~\ref{figfb}
as a function of radius, rescaled by the cosmic value.
None of the analytical cluster models is able to fit satisfactorily the
simulated profile. The canonical \bm is the only scenario in which the
baryon fraction reaches an asymptotic value, but this is due to the
equally wrong gas and dark matter densities. In the other models,
the baryon fraction reaches a maximum around three times the
characteristic radius, but then it drops rather steeply.

We used this maximum to find out the normalisation of the ICM
density, $\rho_0$. However, the steep
asymptotic slope of the gas density profile hints that more realistic
models have to be considered in order to reproduce the
radial structure near the virial radius.
On the other hand, such a modelisation (e.g. including departures from 
spherical symmetry,
infall and turbulent motions) would introduce additional free
parameters that would complicate the physical interpretation of the
results.

 \section{Observable consequences}
 \label{secObs}

Analytical models of galaxy groups and clusters are a key ingredient
in the interpretation of observational data, since they relate the
X-ray emission of the ICM to the underlying gas and dark matter
distributions.
In particular, X-ray observations are usually restricted to the
central regions, and these models allow the extrapolation of the
spherically-averaged profiles up to the virial radius.
In some occasions, they are also extrapolated inwards, in
order to correct for the presence of a central cooling flow.
Quite often, the virial radius itself is derived by assuming a
particular model.

\begin{figure}
  \centering
\includegraphics[width=8cm]{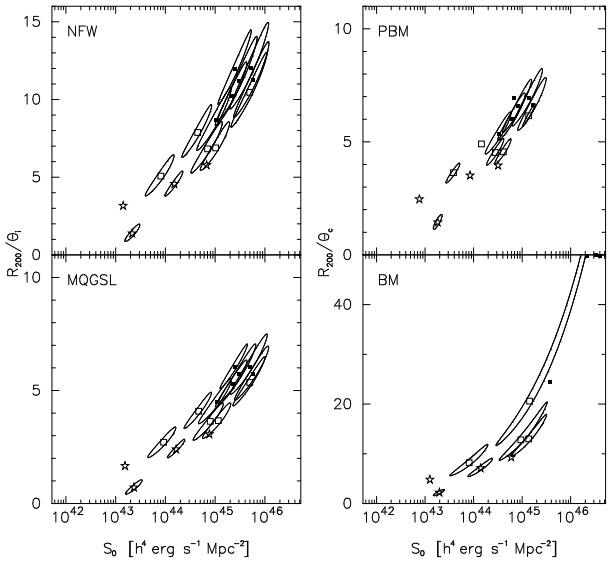}
  \caption{Best-fitting values of the central X-ray surface
    brightness, $S_0$, and the projected characteristic radius,
    $\theta_{\rm i}$.
    Solid squares are used for relaxed systems, empty squares for
    minor mergers and stars for major mergers.
    Contours are drawn at
    $\sqrt{\chi^2/(dof)}=0.2$.
}
  \label{figFitSx}
\end{figure}

Therefore, we would like to investigate in some detail to what extent
the assumed radial structure can bias the conclusions drawn from
observational studies.
We construct X-ray surface brightness profiles in the same manner as we
did for the projected emission-weighted temperature (i.e. assuming
pure bremsstrahlung radiation and averaging over the three axes).
Then, we fit the simulated profiles
with our four analytical models, varying
the projected characteristic radius, $\theta_{\rm i}$,
and the central surface brightness, $S_0$.
Results are shown in Figure~\ref{figFitSx}.

Only data between 0.05 and 0.5 \rv were used.
The lower cut-off is similar to region excised in observational
studies to avoid the cooling flow region.
The upper scale is similar to that attained by X-ray data.
Furthermore, we find that the emission at larger radii is
significantly enhanced by small mass subhaloes, which manifest as
peaks in the surface brightness. We discard these features by fitting
only those bins that lead to a monotonically decreasing profile.
As pointed out by \citet{ME01}, the inclusion of emission lines would
still increase the contribution of substructure to the
total X-ray luminosity.

\begin{figure}
  \centering
\includegraphics[width=8cm]{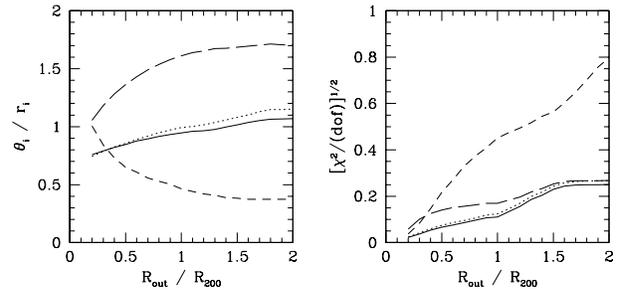}
  \caption{Dependence on the outer radius, $R_{\rm out}$, used for the
    fit to the X-ray surface brightness profile.
    Solid lines represent NFW model; dotted, MQGSL; short-dashed, BM;
    long-dashed, PBM.
    \emph{Left panel:} Projected characteristic radius, divided by the
    three-dimensional estimation based on the CDM/gas distribution.
    \emph{Right panel:} Goodness of the surface brightness fits.
}
  \label{figRout}
\end{figure}

The dependence of the fit on the choice of the outer
radius $R_{\rm out}$ is plotted in Figure~\ref{figRout}.
NFW and MQGSL models are more stable than \BM s, and their
best-fitting $\theta_{\rm i}$ are much closer to the actual
characteristic radius $r_{\rm i}$, computed from the mass
distribution between 0.05 \rv and \rv (see Section~\ref{secResults}).
The conventional \bm offers a poor fit to our simulated surface
brightness at large radii, due to the shallow
slope associated to $\beta=2/3$. For several objects, the best fit is
obtained with the lowest possible $\theta_{\rm c}$ (i.e. a pure power law,
$S_{\rm X}\propto r^{-1.5}$).
Since we assumed $\beta=1$ for the PBM, a better agreement is found
with the numerical results, although the best-fitting $\theta_{\rm c}$
is systematically higher than the core radius inferred from the gas mass.

We assume that the characteristic radii of our clusters in each model
are equal to their best-fitting projected equivalents, $\theta_{\rm i}$.
To obtain the characteristic
CDM densities, as well as the central gas densities and temperatures,
we must solve a system of equations, which in the \bm is given
by (\ref{eqS0}), (\ref{eqRhoBeta}) and (\ref{eqFbBM}).
For the models based on a 'universal' dark matter density profile,
the surface brightness is integrated numerically, yielding
the central values quoted in Table~\ref{tabModels2}.
They must be combined with equations (\ref{eqBi}) and (\ref{eqFbNFW}),
substituting the appropriate value of $B_{\rm i}$.

A last interesting point to note is that we are not using any information
about the X-ray temperature. In doing so, we could skip one of our
constraints, such as the value of the polytropic index or the
normalisation of the central gas density with respect to the CDM component.

 \subsection{X-ray surface brightness}

\begin{figure}
  \centering
\includegraphics[width=8cm]{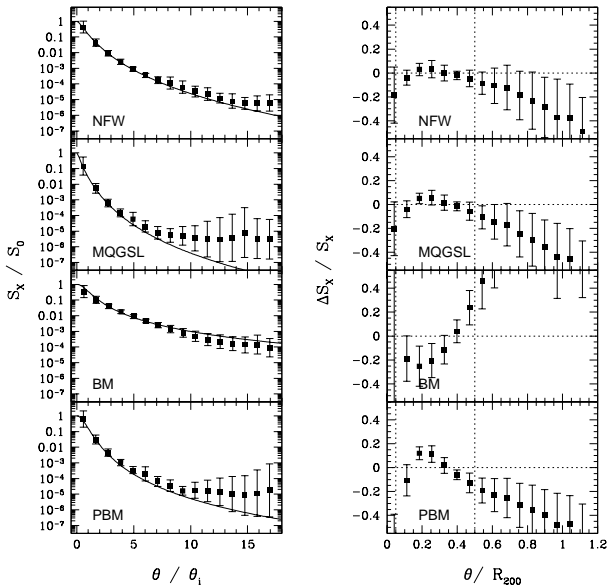}
  \caption{X-ray surface brightness,
    averaged over all clusters except major mergers.
    Error bars indicate one-sigma scatter of individual profiles.
    \emph{Left panel:} Surface brightness profile, scaled by the
    best-fitting $S_0$ and $\theta_{\rm i}$.
    Analytical cluster models are shown as solid lines.
    \emph{Right panel:} Accuracy of each analytical
    profile. Vertical dotted lines mark the fitted region.
}
  \label{figSx}
\end{figure}

The first question we would like to address is how well can our
analytical cluster models fit the X-ray surface brightness.
We show in Figure~\ref{figSx} that the numerical profiles are fairly
described by NFW or MQGSL, although the accuracy of the fits
degrades dramatically at large radii.
Never the less, we find that the
difference between the analytical and simulated profiles is of the
order of $30-40$ per cent up to the virial radius. Some part of it
is caused by the low gas density predicted by these
models at large radii. The rest is due to the emission
of substructure beyond 0.5 \RV.

Consistently with observational results, the isothermal \bm with
$\beta=2/3$ provides a fairly good fit to the X-ray surface
brightness only for $r\le0.3\Rv$. As was shown for the gas density,
the outer parts are better approximated by $\beta\sim1$.
When we let $\beta$ vary as a free parameter, the best fit to our
data is obtained by $\beta\sim0.8$. However, this is a 'compromise
solution' that fails to properly describe the ICM at very
small or large radii.

 \subsection{Mass estimates}

\begin{figure}
  \centering
\includegraphics[width=8cm]{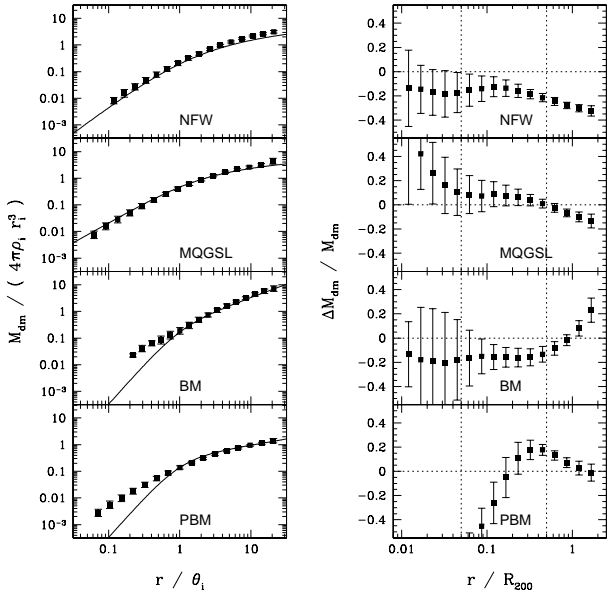}
  \caption{Dark matter mass profile, inferred from the X-ray surface
    brightness. Symbols as in Figure~\ref{figSx}.}
  \label{figM}
\end{figure}

One of the most important applications of hydrostatic equilibrium and
a polytropic equation of state is the estimate of the cumulative
mass profile of both dark and baryonic components,
given the observed X-ray emission.
We plot in Figure~\ref{figM} the CDM mass distribution inferred for our
cluster sample (excluding major mergers), scaled by the characteristic
densities and radii derived from the fit to the surface brightness
profile.

The accuracy of the mass estimate is similar in all models ($\sim20$
per cent errors), but the shape of the mass distribution is substantially
more accurate when a NFW or MQGSL 'universal' CDM density profile is assumed.
The isothermal \bm gives an acceptable estimate of the dark matter
mass for $0.1<r/\Rv<1$ because of the small values inferred for the
core radius.
In the PBM, this quantity is substantially higher and thus a 'core'
in the dark matter distribution can be noticed for $r\le0.3\Rv$.

\begin{figure}
  \centering
\includegraphics[width=8cm]{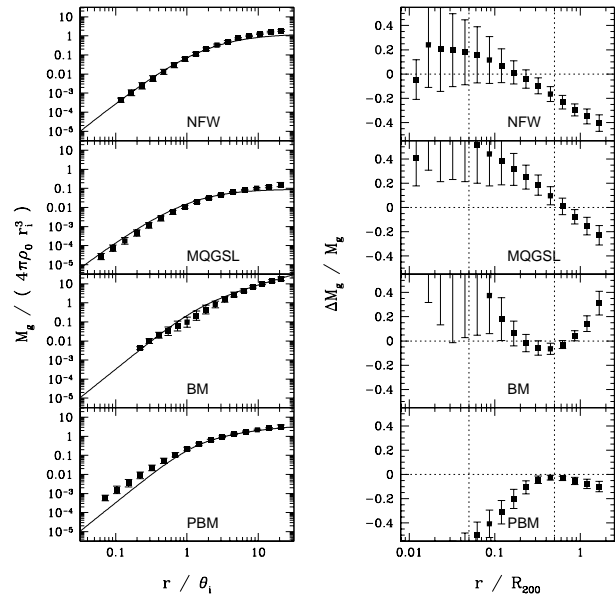}
  \caption{Same as Figure~\ref{figM}, for the cumulative gas mass.}
  \label{figMgas}
\end{figure}

Cumulative gas mass estimates are plotted in Figure~\ref{figMgas}.
In this case, the small values of $r_{\rm c}$ in the BM lead to
extremely high central densities, whereas the PBM (with $\beta=1$)
predicts an excessively large core. Again, a profile closer to the
numerical results can be obtained by treating $\beta$ as a free parameter.
The estimates based on a 'universal' CDM profile are only slightly more
accurate. The scatter, as well as the overall shape, are greatly
improved by considering individual values of the polytropic index.

\begin{figure}
  \centering
\includegraphics[width=8cm]{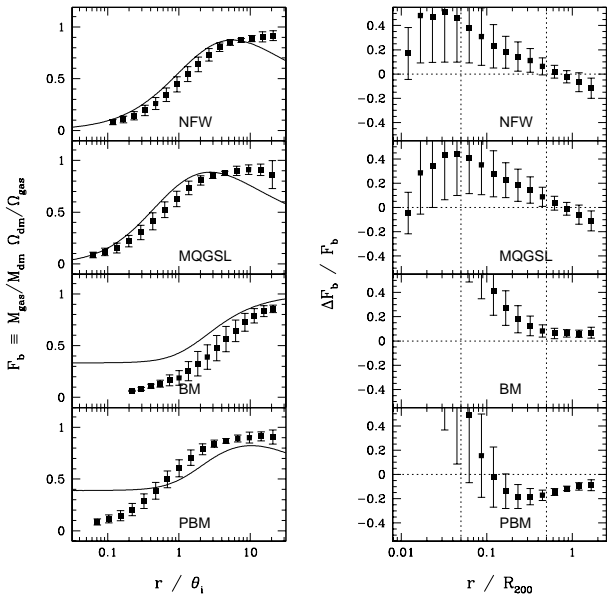}
  \caption{Same as Figure~\ref{figM}, for the cumulative baryon
    fraction in units of the cosmic value.}
  \label{figFB}
\end{figure}

Regarding the total baryon fraction, we can see in Figure~\ref{figFB}
that the \BM s can lead to severe overestimates of this quantity at
small radii ($r\le0.1-0.3\ \Rv$). Near the virial radius, though,
uncertainties are of the order of 10 per cent. NFW and MQGSL
models are accurate within $\sim20$ per cent for $r\ge0.2\Rv$ and
$\sim50$ percent as we move closer to the centre.

Another (sometimes more important) source of uncertainty comes from
the estimation of the virial radius itself. Furthermore, \rv is not a
fixed multiple of the characteristic radius in any analytical model
of galaxy clusters,
and hence the cumulative baryon fraction at a fixed overdensity is
extremely sensitive to halo concentration, particularly at low radii.
Assuming a NFW model, the baryon fraction within \rv can change by
less than 10 per cent from $c=4$ to $c=9$, but at 0.1 \RV, the ratio
of gas to dark matter mass varies from .5 to .75 times the cosmic value.

 \subsection{Entropy}

Entropy plays a fundamental role in clusters of galaxies
because convection acts as an entropy-sorting
device, moving low entropy material to the cluster core and high
entropy material to the outskirts.
Gas density and temperature in hydrostatic and convective equilibrium
are just manifestations of the underlying entropy distribution.
The claim of an entropy floor in small groups \citep{PCN99} has
often been regarded as evidence of preheating of the ICM at high 
redshift.
Using a much larger sample, \citet{0304048} find that the observed
profiles appear to be approximately self-similar apart from a
normalisation constant. They obtain a best-fit behaviour
$S(0.1\Rv)\propto T^{0.65}$ for systems spanning a temperature range
from 1 to 10 keV.

The entropy profile has recently been studied by \citet{Borgani01_sh} and 
\citet{Finoguenov03_sh} by means of numerical simulations that include
cooling and preheating. With gravitational heating alone, these
authors find a power-law entropy profile, consistent with other results
based on the standard implementation of SPH \citep[see e.g.][]{SB_sh}.
As was discussed in Section~\ref{secSchemes}, the situation changes
dramatically when entropy conservation is explicitly enforced, either
in the SPH algorithm or in Eulerian schemes. Shock heating becomes
substantially more efficient, leading to a much higher entropy near
the cluster centre, as well as stripping low-entropy gas from the
infalling galaxies that otherwise would survive for several crossing
times \citep{Borgani01_sh}.

\begin{figure}
  \centering
\includegraphics[width=8cm]{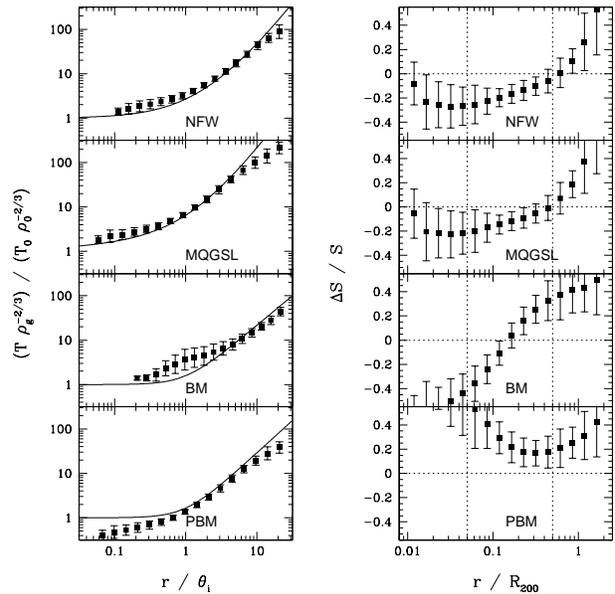}
  \caption{Same as Figure~\ref{figM}, for the entropy profile.
    Entropy is defined as $S=kTn_e^{-2/3}$ and rescaled by
    the central values of the ICM density and temperature inferred
    from the fit to the X-ray surface brightness.}
  \label{figS1}
\end{figure}

We plot our entropy profiles in Figure~\ref{figS1}. As expected
for a polytropic gas in hydrostatic equilibrium, they are not
pure power laws for any reasonable choice of the
'universal' gas or dark matter distribution.
Furthermore, the shape of the entropy profile does not depend
systematically on the mass or temperature of the object,
in agreement with recent observations \citep{0304048}.
NFW or MQGSL models provide a much better estimate of the entropy
distribution than any version of the \BM. However, the low gas densities
predicted at large radii yield a slope at $r\sim\Rv$
significantly steeper than the numerical data.

Given our limited temperature coverage, the normalisation of the
entropy profiles is consistent within the error bars with both the
self-similar scaling ($S\propto T$) as well as with the observed
trend, $S\propto T^{0.65}$.
In any case, we would like to stress that we do not expect to find the 
self-similar scaling albeit we are neglecting non-gravitational
processes. The central entropy in \emph{any} model scales as
$T_0\rho_0^{-2/3}$, where $\rho_0\propto\rhoi$ and
$T_0\propto\rhoi\ri^2$. Recovering a normalisation exactly
proportional to the central temperature would require
that the characteristic CDM density was independent on halo mass
(i.e. constant concentration).

As in the baryon fraction, we argue that the characteristic density
and radius can be more physically meaningful than the mass and radius
at a given overdensity. Depending on concentration, the entropy at
some fraction of \rv can vary as much as a factor of 3.
This can have important consequences on the $S-T$ relation, since
we have shown that merging systems are systematically much less
concentrated than relaxed haloes of the same mass.
A lower concentration implies a lower central gas density,
but also a lower temperature and a smaller value of $\Rv/\ri$.
Therefore, the net effect depends on the details of the
mass-concentration relation.

\begin{figure}
  \centering
\includegraphics[width=8cm]{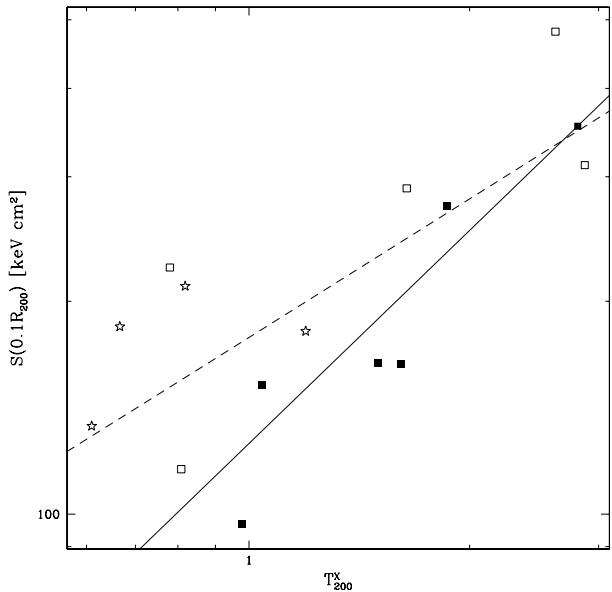}
  \caption{Gas entropy at 0.1 \rv versus emission-weighted
    temperature. Solid squares represent relaxed clusters, empty
    squares are used for minor mergers and stars for major merging
    systems. Solid line depicts the self-similar scaling, $S\propto
    T$, while dashed line shows the observed behaviour, $S\propto T^{0.65}$.
    }
  \label{figS2}
\end{figure}

The $S-T$ relation for our sample of numerical clusters is shown in
Figure~\ref{figS2}. The entropy, $S=kT[\rhogas/(\mu m_{\rm
  p})]^{-2/3}$, is evaluated at 0.1 \rv and plotted as a function of
the X-ray temperature of the object (see Table~\ref{tabSample}).
Systems that have been catalogued as mergers according to our
substructure criterium display higher entropies than relaxed clusters.
Since mergers are more common on group scales, the $S-T$ relation
becomes considerably flattened when these systems are taken into account.

 \section{Conclusions}
 \label{secConclus}

In this paper, we have considered four self-consistent analytical models 
of galaxy clusters, based on the hypotheses that the hot ICM gas is in 
hydrostatic equilibrium with the dark matter halo and that it follows a
polytropic equation of state.
Two of our models assume NFW and MQGSL formulae to describe the CDM
density profile, whereas the other two assume a \bm for the gas
distribution. One is an isothermal version with $\beta=2/3$ (BM) and the
other is a polytropic model with $\gamma=1.18$ and $\beta=1$ (PBM).

We performed a set of high-resolution adiabatic gasdynamical simulations to
assess the accuracy of the basic approximations (i.e. hydrostatic
equilibrium and a polytropic relation).
Then, we compared the
radial distributions of gas and dark matter expected in each
analytical model with the results for a sample of 15 numerical
groups and clusters between 1 and 3 keV.

Our main conclusions can be summarised as follows:

\begin{enumerate}

  \item We find additional evidence of non-physical entropy losses in
    the standard implementation of the SPH algorithm. This effect can
    be of critical importance in the computation of the temperature
    and entropy profiles.
    The new formulation proposed by \citet{gadgetEntro02} appears to
    be a promising alternative to overcome this problem.

  \item Thermally-supported hydrostatic equilibrium is a valid
    approximation for objects classified as 'relaxed' on dynamical
    grounds, up to $\sim0.8\Rv$.
    We find evidence of non-thermal support in merging systems.

  \item A polytropic equation of state with $\gamma\sim1.18$ provides
    a fairly good description of the ICM gas. Substructure induces
    deviations from this relation.

  \item The density profile of our clusters can be well fitted by
    either NFW or MQGSL formulae.
    \BM s predict constant dark matter density near the centre, in
    conflict with the numerical results.
    Major mergers feature a power-law
    density profile with slope $-2\leq\alpha\leq-1$.

  \item Gas temperature declines by a factor of $2-3$ from the centre
    to the virial radius. Our projected X-ray temperature profiles can
    be fitted by the 'universal' form proposed by \citet{Loken02_sh},
    derived from Eulerian gasdynamical simulations,
    although we obtain different values for the free parameters.
    Purely adiabatic physics cannot account for a large isothermal core.

  \item \BM s fail to reproduce the gas density profile. While a fit
    to the inner part yields $\beta\sim2/3$, larger values are
    required at large radii. NFW and MQGSL models are able to predict
    the average distribution, but variations in the polytropic
    index lead to a very large scatter. The asymptotic slope of the
    gas density profile is too steep compared to the simulation data.

  \item Models based on a 'universal' CDM profile are able to estimate
    the ICM properties within $30-40$ per cent errors,
    when applied to the simulated X-ray surface brightness.
    Assuming a \bm yields similar estimates
    for $r\ge0.1\Rv$, but the shape of the inferred profiles at
    smaller radii can be severely misleading.

  \item Contrary to conventional SPH estimates, the entropy
    distribution is not a pure power law.
    All our analytical cluster models predict an increasing slope with 
    radius, consistent with our numerical data.
    The entropy profile is entirely determined by the characteristic
    density and radius of the CDM halo, and the normalisation depends
    on the mass-concentration relation. In particular, it does not
    necessarily scale linearly with gas temperature.

In general terms, we claim that the radial structure of galaxy groups
and clusters can be understood in terms of hydrostatic equilibrium and 
a polytropic equation of state for the gas, at least when only
gravitational processes and adiabatic gasdynamics are considered.

Although \BM s can sometimes provide a fair estimate of several
cluster properties, they fail to provide an overall description that
matches all the numerical data.
Models based on a 'universal' CDM density profile have a similar
or better accuracy, and they are far more reliable in a qualitative
sense (i.e. concerning the shape of the profiles).
We argue that these models are preferable in order to analyse and
interpret observational data.

\end{enumerate}


\section*{Acknowledgements}

We thank Ben Moore for a useful discussion on the CDM density profiles.
We also thank Volker Springel for providing the entropy conserving
version of \G, and Andrey Kravtsov for providing the ART data in
Figure~\ref{figA}.
This work has been partially supported by the MCyT (Spain) under
project number AYA-0973, by the {\em Acciones Integradas
  Hispano-Alemanas} HA2000-0026 and by {\em Deutscher Akademischer
  Austauschdienst} DAAD (Germany). We thank the Ciemat and CEPBA
(Spain) and LRZ (Germany) for allowing us to use their supercomputers
to perform the simulations reported in this paper.


\bibliographystyle{mn2e}
\bibliography{../../BibTeX/DATABASE,../../BibTeX/PREPRINTS,../../BibTeX/SHORT}

\begin{thebibliography}{}

\bibitem[\protect\citeauthoryear{{Ascasibar}}{{Ascasibar}}{2003}]{tesis}
{Ascasibar} Y.,  2003, PhD thesis, Universidad Aut{\' o}noma de Madrid (Spain),
  {\tt(astro-ph/0305250)}

\bibitem[\protect\citeauthoryear{{Bartelmann} \& {Steinmetz}}{{Bartelmann} \&
  {Steinmetz}}{1996}]{BartelmannSteinmetz96}
{Bartelmann} M.,  {Steinmetz} M.,  1996, \mnras, 283, 431

\bibitem[\protect\citeauthoryear{{Bialek}, {Evrard} \& {Mohr}}{{Bialek}
  et~al.}{2001}]{Bialek01}
{Bialek} J.~J.,  {Evrard} A.~E.,    {Mohr} J.~J.,  2001, \apj, 555, 597

\bibitem[\protect\citeauthoryear{{Borgani et~al.}}{{Borgani
  et~al.}}{2001}]{Borgani01_sh}
{Borgani et~al.} 2001, \apjl, 559, L71

\bibitem[\protect\citeauthoryear{{Borgani et~al.}}{{Borgani
  et~al.}}{2002}]{Borgani02_sh}
{Borgani et~al.} 2002, \mnras, 336, 409

\bibitem[\protect\citeauthoryear{{Bryan} \& {Norman}}{{Bryan} \&
  {Norman}}{1995}]{BryanNorman95}
{Bryan} G.~L.,  {Norman} M.~L.,  1995, Bulletin of the American Astronomical
  Society, 27, 1421

\bibitem[\protect\citeauthoryear{{Bryan} \& {Norman}}{{Bryan} \&
  {Norman}}{1998}]{BN98}
{Bryan} G.~L.,  {Norman} M.~L.,  1998, \apj, 495, 80

\bibitem[\protect\citeauthoryear{{Bryan}, {Norman}, {Stone}, {Cen} \&
  {Ostriker}}{{Bryan} et~al.}{1995}]{Bryan95}
{Bryan} G.~L.,  {Norman} M.~L.,  {Stone} J.~M.,  {Cen} R.,    {Ostriker} J.~P.,
   1995, Comput. Phys. Comm., 89, 149

\bibitem[\protect\citeauthoryear{{Burkert} \& {Silk}}{{Burkert} \&
  {Silk}}{1999}]{BurkertSilk99_sh}
{Burkert} A.,  {Silk} J.,  1999, in Dark matter in Astrophysics and Particle
  Physics {\tt(astro-ph/9904159)}

\bibitem[\protect\citeauthoryear{{Cavaliere} \& {Fusco-Femiano}}{{Cavaliere} \&
  {Fusco-Femiano}}{1976}]{CavaliereFusco76}
{Cavaliere} A.,  {Fusco-Femiano} R.,  1976, \aap, 49, 137

\bibitem[\protect\citeauthoryear{{Cen}}{{Cen}}{1992}]{Cen92}
{Cen} R.,  1992, \apjs, 78, 341

\bibitem[\protect\citeauthoryear{{Cen}, {Ostriker}, {Jameson} \& {Liu}}{{Cen}
  et~al.}{1990}]{Cen90}
{Cen} R.~Y.,  {Ostriker} J.~P.,  {Jameson} A.,    {Liu} F.,  1990, \apjl, 362,
  L41

\bibitem[\protect\citeauthoryear{{Col{\'{\i}}n et~al.}}{{Col{\'{\i}}n
  et~al.}}{1999}]{Colin99_sh}
{Col{\'{\i}}n et~al.} 1999, \apj, 523, 32

\bibitem[\protect\citeauthoryear{{Dav{\' e}}, {Katz} \& {Weinberg}}{{Dav{\' e}}
  et~al.}{2002}]{Dave02}
{Dav{\' e}} R.,  {Katz} N.,    {Weinberg} D.~H.,  2002, \apj, 579, 23

\bibitem[\protect\citeauthoryear{{De Grandi} \& {Molendi}}{{De Grandi} \&
  {Molendi}}{2002}]{GrandiMolendi02}
{De Grandi} S.,  {Molendi} S.,  2002, \apj, 567, 163

\bibitem[\protect\citeauthoryear{{Edge} \& {Stewart}}{{Edge} \&
  {Stewart}}{1991}]{EdgeStewart91}
{Edge} A.~C.,  {Stewart} G.~C.,  1991, \mnras, 252, 428

\bibitem[\protect\citeauthoryear{{Eke}, {Navarro} \& {Frenk}}{{Eke}
  et~al.}{1998}]{Eke98}
{Eke} V.~R.,  {Navarro} J.~F.,    {Frenk} C.~S.,  1998, \apj, 503, 569

\bibitem[\protect\citeauthoryear{{Ettori}}{{Ettori}}{2000}]{Ettori00}
{Ettori} S.,  2000, \mnras, 311, 313

\bibitem[\protect\citeauthoryear{{Evrard}, {Metzler} \& {Navarro}}{{Evrard}
  et~al.}{1996}]{EMN96}
{Evrard} A.~E.,  {Metzler} C.~A.,    {Navarro} J.~F.,  1996, \apj, 469, 494

\bibitem[\protect\citeauthoryear{{Finoguenov}, {Arnaud} \&
  {David}}{{Finoguenov} et~al.}{2001}]{Finoguenov01}
{Finoguenov} A.,  {Arnaud} M.,    {David} L.~P.,  2001, \apj, 555, 191

\bibitem[\protect\citeauthoryear{{Finoguenov et~al.}}{{Finoguenov
  et~al.}}{2003}]{Finoguenov03_sh}
{Finoguenov et~al.} 2003, \aap, 398, L35

\bibitem[\protect\citeauthoryear{{Frenk et~al.}}{{Frenk et~al.}}{1999}]{SB_sh}
{Frenk et~al.} 1999, \apj, 525, 554

\bibitem[\protect\citeauthoryear{{Fukushige} \& {Makino}}{{Fukushige} \&
  {Makino}}{1997}]{FukushigeMakino97}
{Fukushige} T.,  {Makino} J.,  1997, \apjl, 477, L9

\bibitem[\protect\citeauthoryear{{Fukushige} \& {Makino}}{{Fukushige} \&
  {Makino}}{2001}]{FukushigeMakino01}
{Fukushige} T.,  {Makino} J.,  2001, \apj, 557, 533

\bibitem[\protect\citeauthoryear{{Ghigna et~al.}}{{Ghigna
  et~al.}}{1998}]{Ghigna98_sh}
{Ghigna et~al.} 1998, \mnras, 300, 146

\bibitem[\protect\citeauthoryear{{Ghigna et~al.}}{{Ghigna
  et~al.}}{2000}]{Ghigna00_sh}
{Ghigna et~al.} 2000, \apj, 544, 616

\bibitem[\protect\citeauthoryear{{Gingold} \& {Monaghan}}{{Gingold} \&
  {Monaghan}}{1977}]{GingoldMonaghan77}
{Gingold} R.~A.,  {Monaghan} J.~J.,  1977, \mnras, 181, 375

\bibitem[\protect\citeauthoryear{{Gnedin}}{{Gnedin}}{1995}]{Gnedin95}
{Gnedin} N.~Y.,  1995, \apjs, 97, 231

\bibitem[\protect\citeauthoryear{{Gottl{\" o}ber}, {Klypin} \&
  {Kravtsov}}{{Gottl{\" o}ber} et~al.}{2001}]{Gottloeber01}
{Gottl{\" o}ber} S.,  {Klypin} A.,    {Kravtsov} A.~V.,  2001, \apj, 546, 223

\bibitem[\protect\citeauthoryear{{Hernquist}}{{Hernquist}}{1993}]{Hernquist93}
{Hernquist} L.,  1993, \apj, 404, 717

\bibitem[\protect\citeauthoryear{{Irwin} \& {Bregman}}{{Irwin} \&
  {Bregman}}{2000}]{IrwinBregman00}
{Irwin} J.~A.,  {Bregman} J.~N.,  2000, \apj, 538, 543

\bibitem[\protect\citeauthoryear{{Jing} \& {Suto}}{{Jing} \&
  {Suto}}{2000}]{JingSuto00}
{Jing} Y.~P.,  {Suto} Y.,  2000, \apjl, 529, L69

\bibitem[\protect\citeauthoryear{{Kaiser}}{{Kaiser}}{1986}]{Kaiser86}
{Kaiser} N.,  1986, \mnras, 222, 323

\bibitem[\protect\citeauthoryear{{Klypin}, {Gottl{\" o}ber}, {Kravtsov} \&
  {Khokhlov}}{{Klypin} et~al.}{1999}]{Klypin99}
{Klypin} A.,  {Gottl{\" o}ber} S.,  {Kravtsov} A.~V.,    {Khokhlov} A.~M.,
  1999, \apj, 516, 530

\bibitem[\protect\citeauthoryear{{Klypin}, {Kravtsov}, {Bullock} \&
  {Primack}}{{Klypin} et~al.}{2001}]{Klypin01}
{Klypin} A.,  {Kravtsov} A.~V.,  {Bullock} J.~S.,    {Primack} J.~R.,  2001,
  \apj, 554, 903

\bibitem[\protect\citeauthoryear{{Komatsu} \& {Seljak}}{{Komatsu} \&
  {Seljak}}{2001}]{KS01}
{Komatsu} E.,  {Seljak} U.,  2001, \mnras, 327, 1353

\bibitem[\protect\citeauthoryear{{Kravtsov}, {Klypin} \& {Hoffman}}{{Kravtsov}
  et~al.}{2002}]{ARThydro02}
{Kravtsov} A.~V.,  {Klypin} A.,    {Hoffman} Y.,  2002, \apj, 571, 563

\bibitem[\protect\citeauthoryear{{Kravtsov}, {Klypin} \& {Khokhlov}}{{Kravtsov}
  et~al.}{1997}]{ART97}
{Kravtsov} A.~V.,  {Klypin} A.~A.,    {Khokhlov} A.~M.,  1997, \apjs, 111, 73

\bibitem[\protect\citeauthoryear{{Lewis et~al.}}{{Lewis
  et~al.}}{2000}]{Lewis00_sh}
{Lewis et~al.} 2000, \apj, 536, 623

\bibitem[\protect\citeauthoryear{{Loken et~al.}}{{Loken
  et~al.}}{2002}]{Loken02_sh}
{Loken et~al.} 2002, \apj, 579, 571

\bibitem[\protect\citeauthoryear{{Lucy}}{{Lucy}}{1977}]{Lucy77}
{Lucy} L.~B.,  1977, \aj, 82, 1013

\bibitem[\protect\citeauthoryear{{Makino}, {Sasaki} \& {Suto}}{{Makino}
  et~al.}{1998}]{Makino98}
{Makino} N.,  {Sasaki} S.,    {Suto} Y.,  1998, \apj, 497, 555

\bibitem[\protect\citeauthoryear{{Markevitch}, {Forman}, {Sarazin} \&
  {Vikhlinin}}{{Markevitch} et~al.}{1998}]{Markevitch98}
{Markevitch} M.,  {Forman} W.~R.,  {Sarazin} C.~L.,    {Vikhlinin} A.,  1998,
  \apj, 503, 77

\bibitem[\protect\citeauthoryear{{Mathiesen} \& {Evrard}}{{Mathiesen} \&
  {Evrard}}{2001}]{ME01}
{Mathiesen} B.~F.,  {Evrard} A.~E.,  2001, \apj, 546, 100

\bibitem[\protect\citeauthoryear{{Moore}, {Governato}, {Quinn}, {Stadel} \&
  {Lake}}{{Moore} et~al.}{1998}]{Moore98}
{Moore} B.,  {Governato} F.,  {Quinn} T.,  {Stadel} J.,    {Lake} G.,  1998,
  \apjl, 499, L5

\bibitem[\protect\citeauthoryear{{Moore}, {Quinn}, {Governato}, {Stadel} \&
  {Lake}}{{Moore} et~al.}{1999}]{Moore99}
{Moore} B.,  {Quinn} T.,  {Governato} F.,  {Stadel} J.,    {Lake} G.,  1999,
  \mnras, 310, 1147

\bibitem[\protect\citeauthoryear{{Motl}, {Burns}, {Loken}, {Norman} \&
  {Bryan}}{{Motl} et~al.}{2003}]{0302427}
{Motl} P.~M.,  {Burns} J.~O.,  {Loken} C.,  {Norman} M.~L.,    {Bryan} G.,
  2003, \apj, {\tt(astro-ph/0302427)}

\bibitem[\protect\citeauthoryear{{Muanwong}, {Thomas}, {Kay} \&
  {Pearce}}{{Muanwong} et~al.}{2002}]{Muanwong02}
{Muanwong} O.,  {Thomas} P.~A.,  {Kay} S.~T.,    {Pearce} F.~R.,  2002, \mnras,
  336, 527

\bibitem[\protect\citeauthoryear{{Nagai} \& {Kravtsov}}{{Nagai} \&
  {Kravtsov}}{2002}]{cluster6}
{Nagai} D.,  {Kravtsov} A.~V.,  2002, \apj, submitted {\tt(astro-ph/0206469)}

\bibitem[\protect\citeauthoryear{{Navarro}, {Frenk} \& {White}}{{Navarro}
  et~al.}{1995}]{NFW95}
{Navarro} J.~F.,  {Frenk} C.~S.,    {White} S.~D.~M.,  1995, \mnras, 275, 720

\bibitem[\protect\citeauthoryear{{Navarro}, {Frenk} \& {White}}{{Navarro}
  et~al.}{1997}]{NFW97}
{Navarro} J.~F.,  {Frenk} C.~S.,    {White} S.~D.~M.,  1997, \apj, 490, 493

\bibitem[\protect\citeauthoryear{{Nelson} \& {Papaloizou}}{{Nelson} \&
  {Papaloizou}}{1993}]{NelsonPapaloizou93}
{Nelson} R.~P.,  {Papaloizou} J.~C.~B.,  1993, \mnras, 265, 905

\bibitem[\protect\citeauthoryear{{Nelson} \& {Papaloizou}}{{Nelson} \&
  {Papaloizou}}{1994}]{NelsonPapaloizou94}
{Nelson} R.~P.,  {Papaloizou} J.~C.~B.,  1994, \mnras, 270, 1

\bibitem[\protect\citeauthoryear{{Neumann} \& {Arnaud}}{{Neumann} \&
  {Arnaud}}{1999}]{NeumannArnaud99}
{Neumann} D.~M.,  {Arnaud} M.,  1999, \aap, 348, 711

\bibitem[\protect\citeauthoryear{{Pearce}, {Thomas}, {Couchman} \&
  {Edge}}{{Pearce} et~al.}{2000}]{Pearce00}
{Pearce} F.~R.,  {Thomas} P.~A.,  {Couchman} H.~M.~P.,    {Edge} A.~C.,  2000,
  \mnras, 317, 1029

\bibitem[\protect\citeauthoryear{{Pen}}{{Pen}}{1995}]{Pen95}
{Pen} U.,  1995, \apjs, 100, 269

\bibitem[\protect\citeauthoryear{{Pen}}{{Pen}}{1998}]{Pen98}
{Pen} U.,  1998, \apjs, 115, 19

\bibitem[\protect\citeauthoryear{{Ponman}, {Cannon} \& {Navarro}}{{Ponman}
  et~al.}{1999}]{PCN99}
{Ponman} T.~J.,  {Cannon} D.~B.,    {Navarro} J.~F.,  1999, \nat, 397, 135

\bibitem[\protect\citeauthoryear{{Ponman}, {Sanderson} \&
  {Finoguenov}}{{Ponman} et~al.}{2003}]{0304048}
{Ponman} T.~J.,  {Sanderson} A.~J.~R.,    {Finoguenov} A.,  2003, \mnras,
  accepted {\tt(astro-ph/0304048)}

\bibitem[\protect\citeauthoryear{{Power et~al.}}{{Power
  et~al.}}{2003}]{Power03_sh}
{Power et~al.} 2003, \mnras, 338, 14

\bibitem[\protect\citeauthoryear{{Rosati}, {Borgani} \& {Norman}}{{Rosati}
  et~al.}{2002}]{Rosati02}
{Rosati} P.,  {Borgani} S.,    {Norman} C.,  2002, \araa, 40, 539

\bibitem[\protect\citeauthoryear{{Sanderson et~al.}}{{Sanderson
  et~al.}}{2003}]{Sanderson03_sh}
{Sanderson et~al.} 2003, \mnras, 340, 989

\bibitem[\protect\citeauthoryear{{Sarazin}}{{Sarazin}}{1986}]{Sarazin86}
{Sarazin} C.~L.,  1986, Reviews of Modern Physics, 58, 1

\bibitem[\protect\citeauthoryear{{Serna}, {Alimi} \& {Chieze}}{{Serna}
  et~al.}{1996}]{Serna96}
{Serna} A.,  {Alimi} J.-M.,    {Chieze} J.-P.,  1996, \apj, 461, 884

\bibitem[\protect\citeauthoryear{{Serna}, {Dom{\'\i}nguez-Tenreiro} \&
  {S{\'a}iz}}{{Serna} et~al.}{2003}]{deva}
{Serna} A.,  {Dom{\'\i}nguez-Tenreiro} R.,    {S{\'a}iz} A.,  2003, \mnras,
  submitted

\bibitem[\protect\citeauthoryear{{Shapiro}, {Martel}, {Villumsen} \&
  {Owen}}{{Shapiro} et~al.}{1996}]{Shapiro96}
{Shapiro} P.~R.,  {Martel} H.,  {Villumsen} J.~V.,    {Owen} J.~M.,  1996,
  \apjs, 103, 269

\bibitem[\protect\citeauthoryear{{Springel} \& {Hernquist}}{{Springel} \&
  {Hernquist}}{2002}]{gadgetEntro02}
{Springel} V.,  {Hernquist} L.,  2002, \mnras, 333, 649

\bibitem[\protect\citeauthoryear{{Springel}, {Yoshida} \& {White}}{{Springel}
  et~al.}{2001}]{gadget01}
{Springel} V.,  {Yoshida} N.,    {White} S.~D.~M.,  2001, New Astronomy, 6, 79

\bibitem[\protect\citeauthoryear{{Suto}, {Sasaki} \& {Makino}}{{Suto}
  et~al.}{1998}]{Suto98}
{Suto} Y.,  {Sasaki} S.,    {Makino} N.,  1998, \apj, 509, 544

\bibitem[\protect\citeauthoryear{{Tornatore et~al.}}{{Tornatore
  et~al.}}{2003}]{03T_sh}
{Tornatore et~al.} 2003, \mnras, {\tt(astro-ph/0302575)}

\bibitem[\protect\citeauthoryear{{Voit}, {Balogh}, {Bower}, {Lacey} \&
  {Bryan}}{{Voit} et~al.}{2003}]{0304447}
{Voit} G.~M.,  {Balogh} M.~L.,  {Bower} R.~G.,  {Lacey} C.~G.,    {Bryan}
  G.~L.,  2003, \apj, in press {\tt(astro-ph/0304447)}

\bibitem[\protect\citeauthoryear{{White}}{{White}}{2000}]{White00}
{White} D.~A.,  2000, \mnras, 312, 663

\end{thebibliography}

\appendix

\section{Cluster Atlas}
\label{secAtlas}

In the following image, we display our 15 simulated clusters,
projected along the three major axes.
Colours indicate the dark matter surface density along the line of
sight. Contour plots represent the X-ray emission in conventional units.

Quite interestingly, it can be appreciated that many images of merging
systems (both minor and major) appear to be spherically symmetric due
to projection effects.
Usually, a lot more substructure is present in the CDM density (which
roughly corresponds to the optical emission) than in the hot ICM gas
(X-ray).

Indeed, we find that the gaseous component of infalling
galaxies is very efficiently stripped, while dark matter haloes are
able to survive during many orbits around the cluster centre
\footnote{
A visual impression of this phenomenon can be obtained from a series
of animations that can be accessed at
{\tt http://pollux.ft.uam.es/gustavo/VIDEOS/LCDM80/clustervideo.html}
}.
We expect these haloes to retain some of their original gas when
cooling is taken into account, probably undergoing an intense star
formation burst. However, all gas in a warm-hot reservoir is lost
during the first orbit, quenching any subsequent star formation activity.
Numerical experiments that include self-consistent cooling, star
formation and feedback are needed in order to make a quantitative
assessment of this scenario.

\newpage

  \includegraphics{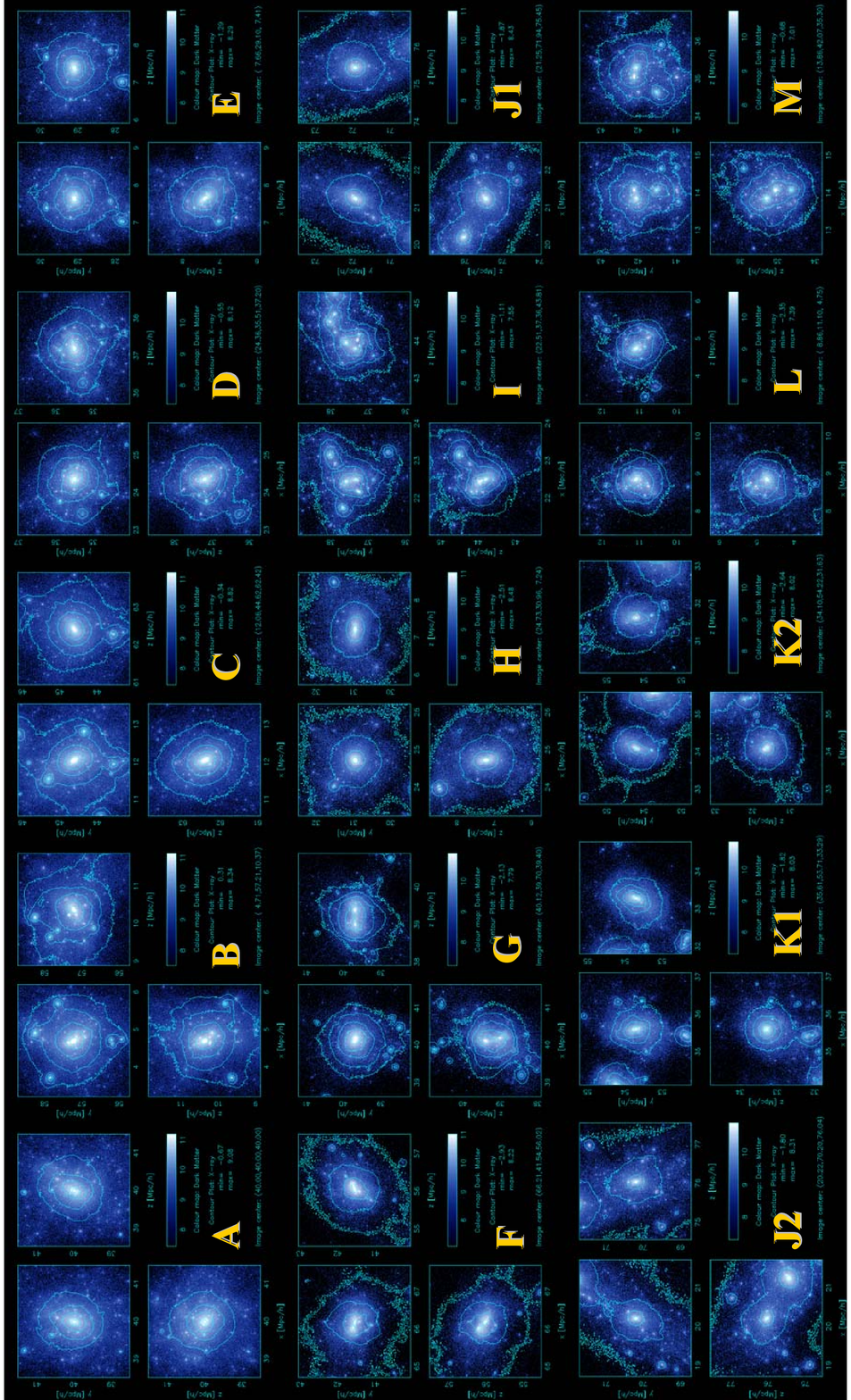}
                  
\end{document}